\providecommand{\bibinfo}[2]{#2}
\documentclass[raggedbottom]{aastex701}
\usepackage{amsmath,amssymb,booktabs,placeins}
\usepackage{etoolbox}
\AtBeginDocument{%
  \setlength{\textfloatsep}{4pt}%
  \setlength{\floatsep}{4pt}%
  \setlength{\intextsep}{4pt}%
  \setlength{\abovecaptionskip}{2.5pt}%
  \setlength{\belowcaptionskip}{1.5pt}%
  \setlength{\abovedisplayskip}{5pt plus 2pt minus 1pt}%
  \setlength{\belowdisplayskip}{5pt plus 2pt minus 1pt}%
  \setlength{\abovedisplayshortskip}{2pt plus 1pt}%
  \setlength{\belowdisplayshortskip}{3pt plus 1pt}%

}
\makeatletter
\patchcmd{\section}{9pt plus 1pt minus 1pt}{6pt plus 1pt minus 1pt}{}{}
\patchcmd{\subsection}{9pt plus 1pt minus 1pt}{5pt plus 1pt minus 1pt}{}{}
\patchcmd{\section}{\center}{\centering}{}{}
\patchcmd{\subsection}{\center}{\centering}{}{}
\patchcmd{\subsection}{{4pt}}{{\baselineskip}}{}{\PackageError{N2}{Subsection spacing patch failed}{Check the AASTeX subsection definition.}}
\patchcmd{\thebibliography}{\vspace{20pt}}{\vspace{7pt}}{}{}
\makeatother
\hypersetup{colorlinks=true,citecolor=blue,urlcolor=blue,linkcolor=blue}
\shorttitle{Active asteroid P/2026 N2}
\shortauthors{Bolin et al.}
\newcommand{\af}{Af\rho}
\newcommand{\obj}{P/2026 N2}
\journalinfo{Accepted for publication in The Astrophysical Journal Letters, 2026 December 2}
\begin{document}
\title{Vera C. Rubin Observatory, Keck, and Gemini characterization of probable Main Belt comet P/2026~N2}

\correspondingauthor{Bryce T. Bolin}
\author[orcid=0000-0002-4950-6323]{Bryce T. Bolin}
\affiliation{Eureka Scientific, Inc., 2452 Delmer Street, Suite 100, Oakland, CA 94602-3017, USA}
\email[show]{bolin.astro@gmail.com}

\author[orcid=0000-0002-2250-8687]{Leo Y. Alcorn}
\affiliation{W. M. Keck Observatory, 65-1120 Mamalahoa Highway, Kamuela, HI 96743, USA}
\email{lalcorn@keck.hawaii.edu}

\author{Leila Alamos}
\affiliation{International Gemini Observatory/NSF NOIRLab, 670 N. A`ohoku Place, Hilo, HI 96720, USA}
\email{leila.alamos@noirlab.edu}

\author[orcid=0009-0008-9808-0411]{Max Brodheim}
\affiliation{W. M. Keck Observatory, 65-1120 Mamalahoa Highway, Kamuela, HI 96743, USA}
\email{}

\author[orcid=0000-0002-9020-5004]{Kristin Chiboucas}
\affiliation{International Gemini Observatory/NSF NOIRLab, 670 N. A`ohoku Place, Hilo, HI 96720, USA}
\email{kristin.chiboucas@noirlab.edu}

\author[orcid=0000-0002-6822-2254]{Emanuele Paolo Farina}
\affiliation{International Gemini Observatory/NSF NOIRLab, 670 N. A`ohoku Place, Hilo, HI 96720, USA}
\email{emanuele.farina@noirlab.edu}

\author[orcid=0000-0002-5025-4645]{Alexa Gordon}
\affiliation{Department of Astronomy, California Institute of Technology, MC 249-17, 1200 East California Boulevard, Pasadena, CA 91125, USA}
\email{}

\author[orcid=0009-0007-5296-4046]{Lordrick Kahinga}
\affiliation{Department of Astronomy and Astrophysics, University of California, Santa Cruz, 1156 High Street, Santa Cruz, CA 95064, USA}
\email{lkahinga@ucsc.edu}

\author[orcid=0000-0002-3370-187X]{Eunchong Kim}
\affiliation{International Gemini Observatory/NSF NOIRLab, 670 N. A`ohoku Place, Hilo, HI 96720, USA}
\email{eunchong.kim@noirlab.edu}

\author[orcid=0000-0002-1428-7036]{Brian C. Lemaux}
\affiliation{International Gemini Observatory/NSF NOIRLab, 670 N. A`ohoku Place, Hilo, HI 96720, USA}
\affiliation{Department of Physics and Astronomy, University of California, Davis, 1 Shields Avenue, Davis, CA 95616, USA}
\email{brian.lemaux@noirlab.edu}

\author[orcid=0000-0001-9589-3793]{Michael Lundquist}
\affiliation{W. M. Keck Observatory, 65-1120 Mamalahoa Highway, Kamuela, HI 96743, USA}
\email{mlundquist@keck.hawaii.edu}

\author{Rita Morris}
\affiliation{W. M. Keck Observatory, 65-1120 Mamalahoa Highway, Kamuela, HI 96743, USA}
\email{}

\author[orcid=0000-0002-7738-6875]{J. Xavier Prochaska}
\affiliation{Department of Astronomy and Astrophysics, University of California, Santa Cruz, 1156 High Street, Santa Cruz, CA 95064, USA}
\email{jxp@ucsc.edu}

\author[orcid=0000-0002-2536-1633]{Hyewon Suh}
\affiliation{International Gemini Observatory/NSF NOIRLab, 670 N. A`ohoku Place, Hilo, HI 96720, USA}
\email{hyewon.suh@noirlab.edu}

\begin{abstract}
Images from 2026 July 6 revealed P/2026 N2, the first comet discovery credited to Vera C. Rubin Observatory. Pre-survey (June 27), survey (July), and Keck (October 7) images resolve this active asteroid's dust across perihelion over 102 days. The geometry- and phase-corrected $i$-band brightness changed by $0.07\pm0.07$ mag within $\sim$1,000 km; the October-minus-July $r$ difference was $-0.10\pm0.09$ mag within 2,000 km. After a single zero-speed June 27 release, radiation pressure would sweep grains of radius $\lesssim1.7$~mm beyond this aperture by October. Keck photometry requires surviving grains to dominate July's scattering area, whereas tail models for other outer Main Belt comets (e.g., 313P and 288P) infer smaller grains. Keck colors ($g-r=0.50\pm0.05$, $r-i=0.06\pm0.06$ mag) indicate nearly neutral reflectance; $A(0^\circ)f\rho=3.66\pm0.13$ cm in $r$ and $3.46\pm0.17$ cm in $i$. Dark Energy Camera images recover N2 (2021 July 4) without establishing activity. Gemini North imaged a tentative compact counterpart on November 3 ($J$) and 20 ($J,Y$; apparent $Y-J=+0.29^{+0.41}_{-0.48}$~mag). Integrated forward 50 Myr, 5,000 clones (nominal $a=3.07$ au, $e=0.21$, $i=2.25^\circ$) retain a concentrated outer Main Belt core. N2's orbit, B/C-like reflectance, and activity across perihelion favor sublimation in this probable Main Belt comet. Perihelion recurrence would test this interpretation.

\end{abstract}

\section{Introduction}
Dust in the Main Belt can be released by ice sublimation, an impact, or rotational mass shedding \citep{Hsieh2006,Jewitt2012}. We use active asteroid for an object with asteroid-like dynamics and observed mass loss, and Main Belt comet (MBC) for a Main Belt object whose activity is attributed to sublimation \citep{Jewitt2012,Hsieh2014}. Recurrence near perihelion supports sublimation in 313P/Gibbs \citep{Hsieh2015a} and 433P/(248370) 2005 QN$_{173}$ \citep{Chandler2021QN173}; water was detected directly in 238P/Read \citep{Kelley2023} and 358P/PANSTARRS \citep{Hsieh2025}. We test N2's classification as a Main Belt comet using its orbit, optical colors, and activity across perihelion. We also assess whether a single dust release can account for its July and October brightness and search for a compact near-infrared counterpart in November.

Madeleine K. McLeod discovered \obj{} in the 2026 July 6 Legacy Survey of Space and Time (LSST) images taken with the Simonyi Survey Telescope at NSF--DOE Vera C. Rubin Observatory\footnote{NSF and DOE denote the U.S. National Science Foundation and Department of Energy, respectively.} and reported it to the Minor Planet Center (MPC) on July 17 \citep{McLeod2026N2CBET}.\footnote{McLeod-LSST is a possible name recognizing the discoverer and survey. We use the designation in the discovery circulars.} McLeod and A. Fitzsimmons confirmed activity at the New Technology Telescope on July 13. C. O. Chandler and collaborators confirmed activity at Apache Point Observatory on July 16 and submitted additional Simonyi measurements \citep{McLeod2026N2MPEC}. R. Weryk independently identified N2 in July 15 images from the first Panoramic Survey Telescope and Rapid Response System telescope (Pan-STARRS1) \citep{McLeod2026N2CBET}. The Minor Planet Electronic Circular (MPEC) assigns discovery credit to the July 6 Simonyi observation (X05), subject to assessment of earlier observations. This is the first comet discovery credited to LSST \citep{McLeod2026N2MPEC}.\footnote{See the \href{https://sbnmpc.astro.umd.edu/mpecwatch/byStation/station_X05.html}{MPEC Watch discovery record for station X05}.} LSST began on the night of 2026 June 29; only the June 27 images reported here are pre-LSST.\footnote{See the \href{https://survey-strategy.lsst.io/progress/}{Rubin Observatory survey status record}.}

The Jet Propulsion Laboratory (JPL) Horizons orbit \citep{Giorgini1996} places N2 in the outer Main Belt ($q=2.42$ au, $P=5.39$ yr; Table~\ref{tab:orbit}). N2 passed perihelion around 2026 September 25--26. We search archival images for N2 at earlier apparitions and integrate orbital clones to test the stability of its present orbit.

\section{Observations and Analysis}
\subsection{Simonyi Survey Telescope imaging}\label{sec:obs_prelsst}
We measured N2 on the Simonyi science cutouts obtained through the Automatic Learning for the Rapid Classification of Events (ALeRCE) broker \citep{Rubin2026Alerts}. The LSST Camera (LSSTCam) recorded the images, which were processed by the Rubin Science Pipelines \citep{LSSTCam2025,RubinSciencePipelines2025,Rubin2026AlertProduction}. The data comprise $i$-band images from June 27 (pre-LSST) and July 6 and 9 (LSST), and $r$-band LSST images from July 9 and July 14. We give observing dates in Universal Time (UT) and clock times in Coordinated Universal Time (UTC). Appendix~\ref{app:rubin} describes the alert products, observing times, and point-spread function (PSF) measurements.

\subsection{Keck I optical imaging}\label{sec:obs_keck}
We obtained Keck I follow-up imaging with the Low Resolution Imaging Spectrometer \citep[LRIS;][]{Oke1995} on UT 2026 October 7 in $g$, $R$, and $i$, under target-of-opportunity program K450 (principal investigator L. Alcorn), with L. Alcorn and B. Bolin as observers. We used $R$ because it was installed in the red-camera filter wheel and $r$ was unavailable. The five $g$, two $R$, and three $i$ exposures total 600, 600, and 360 s, respectively. Appendix~\ref{app:keck} describes the LRIS reduction and calibration to Sloan Digital Sky Survey (SDSS) $g/i$ and Cousins $R_C$.

Keck exposures span 05:22:13--05:46:30 UTC at airmass 1.37--1.38. With $2\times2$ binning, the pixel scales are $0.2714''$ in $g$ and $0.27''$ in $R/i$. Stellar full widths at half maximum (FWHMs) in sidereal reference frames are $0.72''$ in $g$ and $0.78''$ in $R$. Table~\ref{tab:photo} lists the exposure-weighted Keck epochs.

Following \citet{Bolin20202I}, we compare optical fluxes within a projected radius of approximately 1,000 km: $0.96''$ in the alert cutouts and $0.6914''$ in each Keck band. The cutouts limit the available sky annuli (Appendix~\ref{app:phot}; Section~\ref{sec:brightness}).

\subsection{Gemini North imaging}
On UT 2026 November 3 we observed the field of N2 with the Gemini Near-Infrared Spectrograph (GNIRS; \citealt{Elias2006GNIRS}) at Gemini North in short-camera $J$-band imaging mode (JPHOT) (program GN-2026B-FT-207). Twilight saturated the first two of 17 science frames. The remaining 15 frames (two 30~s coadds each) total 900~s at mean epoch 05:00:49.767 UTC. We used the telescope dithers to measure the detector scale and orientation and a cataloged field star to set the origin of the world coordinate system (WCS). We masked the star as it passed near N2's predicted position and fit a compact-source profile to the individual exposures (Appendix~\ref{app:gnirs}). We report $J$ magnitudes on the Mauna Kea Observatories (MKO) Vega system.

We returned to N2 with GNIRS on November 20 under the same program. We obtained 17 $J$ exposures of 60~s each (1,020~s; 04:29:38--04:55:27 UTC), followed by 10 $Y$ exposures of 120~s each (1,200~s; 04:57:53--05:26:50 UTC). We registered the dithered images to cataloged field sources. The brightest $Y$ peak within the combined Horizons and WCS $3\sigma$ ellipse fixed the measurement position in both bands, $0.45''$ west and $0.30''$ south of the prediction. We calibrated both bands with five same-night observations per filter of the standard GSPC S813-D; $J$ is on the MKO Vega system and $Y$ is tied to the Vega standard of \citet{Hillenbrand2002Y}. Appendix~\ref{app:gnirs} describes the calibration and compact-source fits.

\begin{table}[!htbp]
\centering
\caption{Optical aperture photometry, tentative near-infrared compact-source photometry, and observing geometry.}\label{tab:photo}
\small\setlength{\tabcolsep}{2.4pt}
\begingroup\let\tabular\savetabular
\begin{tabular}{llccrrrrrr}
\toprule
Facility & \shortstack{UT date/time\\(2026)} & Band & System & \shortstack{Helio.\\$r_h$ (au)} & \shortstack{Geo.\\$\Delta_{\rm geo}$ (au)} & \shortstack{Phase\\$\alpha$ ($^\circ$)} & \shortstack{Orbit plane\\$\theta$ ($^\circ$)} & $m_\lambda$ (mag) & $H_{\rho,\lambda}$ (mag)\\
\midrule
Rubin & \shortstack{Jun 27\\07:42:34} & $i$ & AB & 2.465 & 1.500 & 9.559 & -0.025 & $22.31\pm0.09$ & $19.08\pm0.09$\\
Rubin & \shortstack{Jul 6\\06:46:09} & $i$ & AB & 2.457 & 1.458 & 5.678 & +0.207 & $22.12\pm0.06$ & $19.10\pm0.06$\\
Rubin & \shortstack{Jul 9\\05:38:58} & $r$ & AB & 2.455 & 1.448 & 4.332 & +0.285 & $22.08\pm0.05$ & $19.14\pm0.05$\\
Rubin & \shortstack{Jul 9\\06:17:21} & $i$ & AB & 2.455 & 1.448 & 4.320 & +0.286 & $21.98\pm0.05$ & $19.04\pm0.05$\\
Rubin & \shortstack{Jul 14\\05:47:37} & $r$ & AB & 2.451 & 1.436 & 2.022 & +0.417 & $21.89\pm0.03$ & $19.07\pm0.03$\\
Keck & \shortstack{Oct 7\\05:37:48} & $g$ & AB & 2.423 & 2.024 & 23.886 & +1.102 & $23.81\pm0.03$ & $19.53\pm0.03$\\
Keck & \shortstack{Oct 7\\05:40:30} & $R_C$ & Vega & 2.423 & 2.024 & 23.886 & +1.101 & $23.12\pm0.04$ & $18.84\pm0.04$\\
Keck & \shortstack{Oct 7\\05:25:36} & $i$ & AB & 2.423 & 2.024 & 23.886 & +1.102 & $23.25\pm0.05$ & $18.97\pm0.05$\\
Gemini N. & \shortstack{Nov 3\\05:00:50} & $J$ & MKO Vega & 2.430 & 2.361 & 23.845 & $+0.783$ & $22.50^{+0.36}_{-0.27}$ & \nodata\\
Gemini N. & \shortstack{Nov 20\\04:42:28} & $J$ & MKO Vega & 2.438 & 2.573 & 22.538 & $+0.541$ & $23.15^{+0.44}_{-0.31}$ & \nodata\\
Gemini N. & \shortstack{Nov 20\\05:12:21} & $Y$ & Vega     & 2.438 & 2.573 & 22.536 & $+0.541$ & $23.43^{+0.27}_{-0.22}$ & \nodata\\
\bottomrule
\end{tabular}\par\endgroup
\par\vspace{2pt}\begin{minipage}{0.98\textwidth}\footnotesize
$r_h$ and $\Delta_{\rm geo}$ are the heliocentric and geocentric distances; $\alpha$ is the solar phase angle; $\theta$ is Earth's signed elevation above the comet's orbital plane, positive toward the reference-frame $+z$ direction. The June 27 Simonyi observation is pre-LSST; the July observations are LSST. Times are UTC observation epochs for the Rubin images and exposure-weighted midpoints for the Keck and Gemini stacks, rounded to the nearest second. Geometry is from JPL Horizons at each exposure or stack mean epoch. Absolute aperture magnitudes use Equation~\ref{eq:h}, the original observing-site geometry, and the zero-phase Halley--Marcus dust correction. Uncertainties are photometric and exclude phase-law systematics. Aperture radii are $0.96''$ for Rubin and $\simeq0.6914''$ for Keck, both corresponding to $\sim$1,000 km at the respective comet distances. The optical measurements include unresolved nucleus light and are uncorrected for PSF-dependent aperture losses. Keck $g/i$ are calibrated to SDSS using the Pan-STARRS1 catalog and the filter transforms of \citet{Tonry2012}; Rubin measurements retain their native $r/i$ passbands. Keck $R_C$ is Cousins $R$ on the Vega system. AB denotes the constant-flux-density magnitude system; Vega magnitudes use Vega as the reference. The Gemini rows give tentative compact-source magnitudes from native-frame fits at the November 3 image-selected position and the November 20 $Y$-selected position. Their pointwise signal-to-noise ratios (S/N) are 3.5, 3.02, and 4.55, respectively. Magnitude errors are statistical; separate calibration floors are 0.06, 0.026, and 0.040 mag. November 20 $Y$ is tied to the Hillenbrand Vega system. At the nominal November 3 position, $J_{\rm MKO,Vega}>22.28$ mag. The listed Gemini magnitudes are apparent magnitudes at the observed epochs. The two-night $J$ comparison is scaled to the November 20 geometry (Appendix~\ref{app:gnirs}). Gemini viewing geometry is geocentric (Horizons solution~3); the plate predictions use Maunakea.
\end{minipage}
\end{table}

\begin{figure}[!htbp]
\centering
\includegraphics[trim=0bp 5bp 0bp 8bp,clip,width=\textwidth]{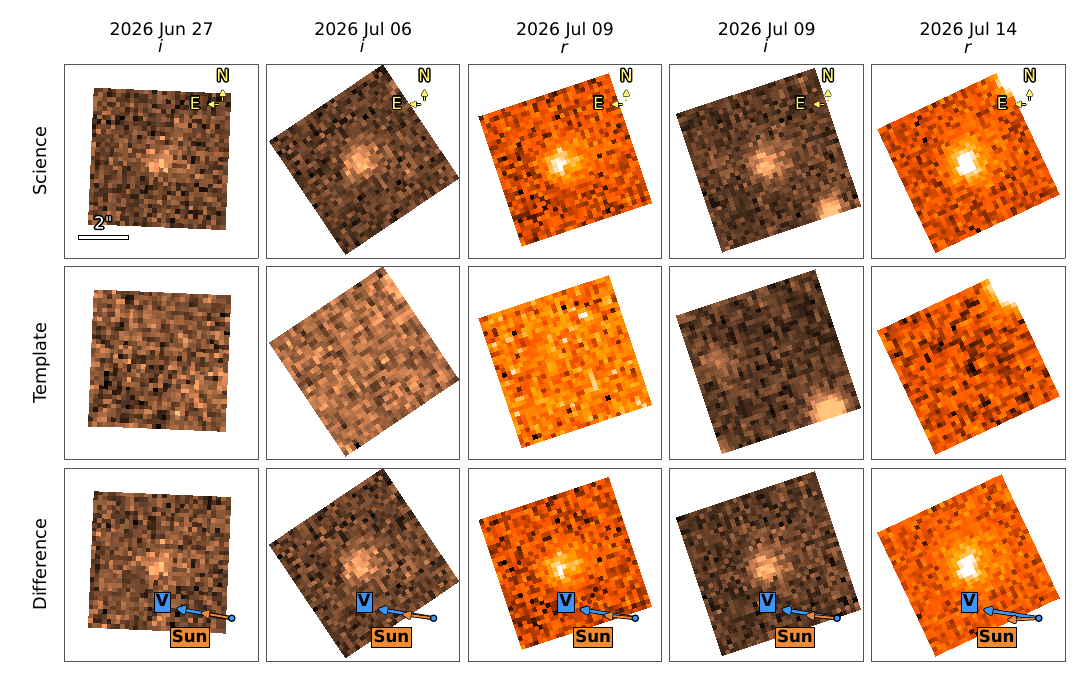}
\caption{Simonyi science (top), reference-template (middle), and difference (bottom) cutouts of P/2026 N2 from 2026 June 27--July 14, labeled by UT date and band. The first column shows pre-LSST data; the four July columns show LSST data. Each $8.5''$ square panel is north up and east left. Affine transforms from the headers place the native $0.20''$ pixels without smoothing, array resampling, or cropping; white regions contain no data. Panels use independent linear ZScale stretches, with $r/i$ colors matching LRIS $R/i$, respectively. The white bar marks $2''$, and the comet is centered in each science cutout. Bottom-row arrows point toward the Sun (orange, Sun) and along the heliocentric velocity (blue, V), calculated from JPL Horizons at each epoch. Each pair shares a blue circular base; arrow lengths are arbitrary.}\label{fig:rubin}
\end{figure}

\begin{figure}[!htbp]
\centering
\includegraphics[width=0.92\textwidth]{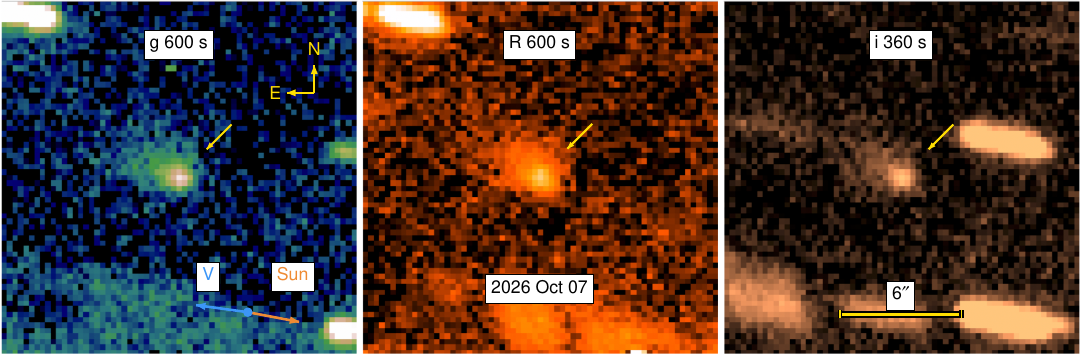}\par\vspace{2pt}
\includegraphics[trim=0bp 2bp 0bp 2bp,clip,width=0.92\textwidth]{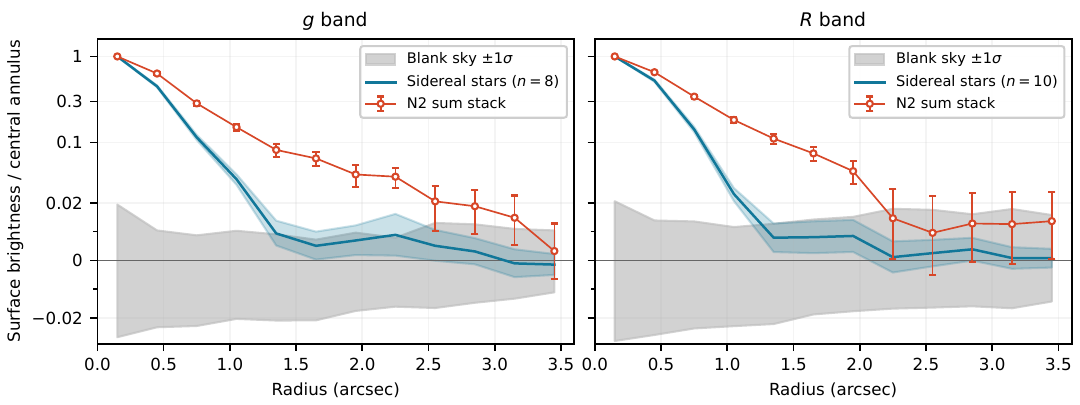}
\caption{Keck/LRIS images and radial profiles of N2 on UT 2026 October 7. Top: $g$, $R$, and $i$ summed images with integrations of 600, 600, and 360 s, respectively. Small yellow arrows identify N2; north is up and east left. The orange Sun and blue V arrows point toward the Sun and along the heliocentric velocity; the outlined bar marks $6''$. Bottom: profiles measured in $0.3''$ annuli and normalized to the central annulus. Red points show N2; blue curves show the mean profiles of eight $g$ and ten $R$ stars in sidereal reference exposures, with bootstrap 16th--84th percentiles shaded. The gray band gives the median and $1\sigma$ scatter at blank positions. Red error bars propagate blank-position covariance and source counting noise through the normalization. Lines connect the measurements without a fitted radial function. The ordinate is logarithmic outside $\pm0.02$ and linear near zero, retaining negative sky-subtracted noise.}\label{fig:psf}
\end{figure}

\begin{figure}[!htbp]
\centering
\includegraphics[width=0.81\textwidth,height=0.78\textheight,keepaspectratio]{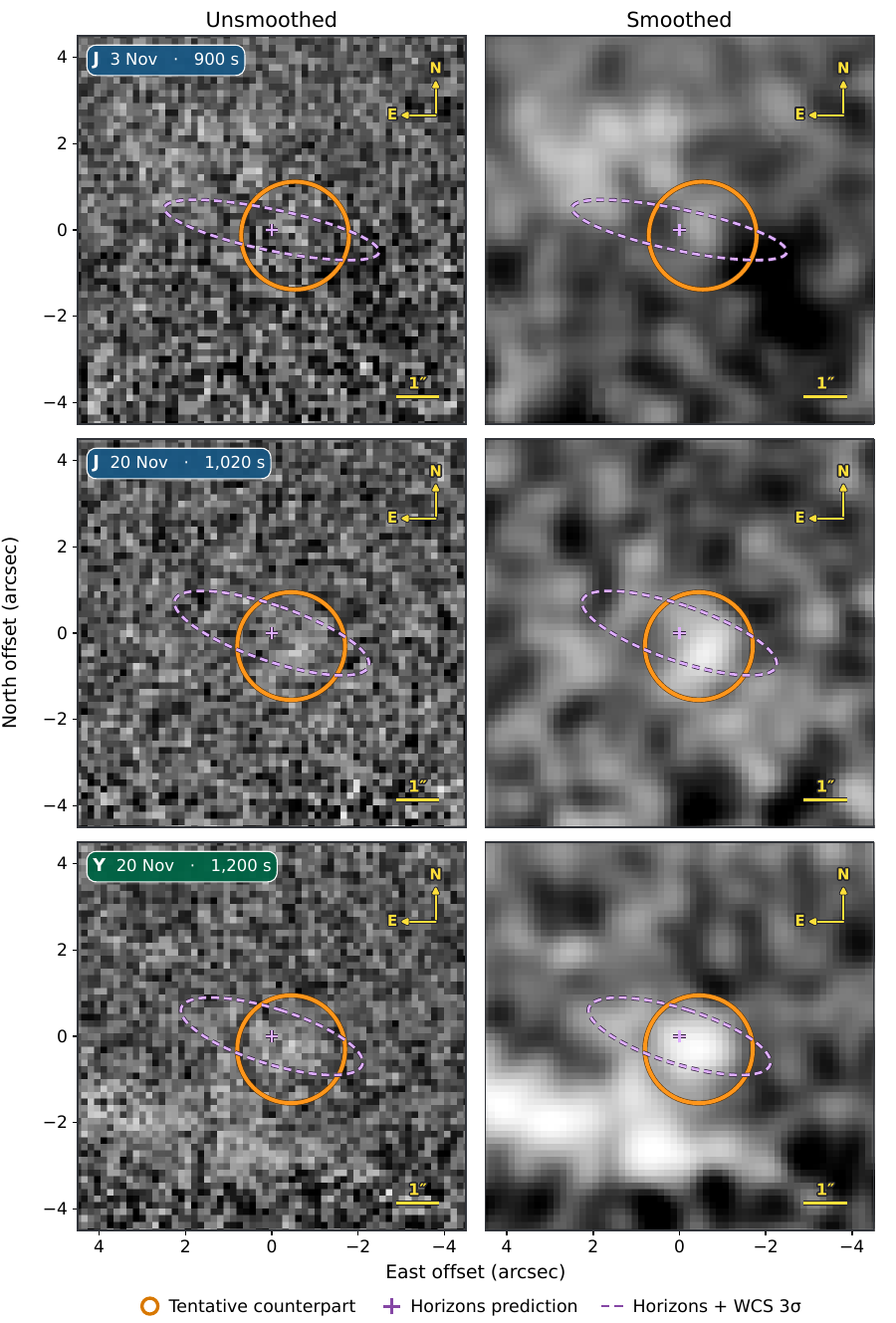}
\caption{Gemini North/GNIRS images of the field of P/2026 N2. From top to bottom, the stacks show November~3 $J$ (900~s), November~20 $J$ (1,020~s), and November~20 $Y$ (1,200~s); the two $J$ nights total 1,920~s. Each row shows the same $9''\times9''$ field unsmoothed (left) and with Gaussian smoothing for display ($\sigma=0.30''$; right). The grayscale stretch is relative to the sky noise in each panel. North is up, east is left, and the yellow bars span $1''$. Offsets are from the Horizons prediction for each stack (purple cross); the dashed purple ellipse encloses the combined Horizons and WCS $3\sigma$ positional uncertainty. Orange circles of radius $1.25''$ mark the tentative counterpart: the feature selected in $J$ on November~3, and the position selected in $Y$ on November~20, $0.45''$ west and $0.30''$ south of the prediction in both bands. Table~\ref{tab:photo} gives the photometry from fits to the individual unsmoothed exposures. The diffuse patch northeast of the November~3 feature is unidentified.}\label{fig:gnirs}
\end{figure}

Figures~\ref{fig:rubin}, \ref{fig:psf}, and \ref{fig:gnirs} show the Simonyi cutouts, Keck images and radial profiles, and Gemini North $J$ and $Y$ images, respectively.

\subsection{Archival imaging and orbital phase}
Pan-STARRS1 warps from the Space Telescope Science Institute image service show no convincing detection at N2's predicted position on 2010 August 5, 6, or 30. We identified the 2021 July 4 Dark Energy Camera (DECam) sequence in the Canadian Astronomy Data Centre Solar System Object Image Search \citep{Gwyn2012SSOS} and retrieved the calibrated exposures from the National Optical-Infrared Astronomy Research Laboratory (NOIRLab) Astro Data Archive. With the same service and the 2026 November 3 Horizons solution, we searched the Canada--France--Hawaii Telescope (CFHT) archive from 1990 January 1 through 2026 November 8 and checked whether the predicted positions fell on the detectors (Appendix~\ref{app:cfht}). We calculate osculating true anomalies and locate earlier perihelia along the perturbed Horizons trajectory (Section~\ref{sec:orbit}).

In the DECam images we test for extension perpendicular to N2's motion \citep{Hsieh2014} using flux-matched stellar PSFs \citep{Hsieh2015}.

\subsection{Colors and relative reflectance}
We derive Keck $r$ from $g$ and Cousins $R_C$ using the Lupton stellar transformation (Equation~\ref{eq:rtransform}). We retain the resulting color covariance when comparing the measurements with the 24 Bus--DeMeo class means \citep{DeMeo2009}. The LSST July 9 $r-i$ color provides an independent comparison (Appendix~\ref{app:phot}).

\subsection{Dust and brightness measurements}
We compare epochs using the phase-corrected absolute aperture magnitude
\begin{equation}
H_{\rho,\lambda}=m_{\rho,\lambda}-5\log_{10}(r_h\Delta)
+2.5\log_{10}\Phi(\alpha),
\label{eq:h}
\end{equation}
Here $r_h$ and $\Delta$ are the numerical heliocentric and observer distances in au, $\alpha$ is the phase angle, and $\Phi(0)=1$. We adopt the zero-phase-normalized Halley--Marcus composite dust phase function tabulated by Lowell Observatory.\footnote{\url{https://comet.lowell.edu/comet/dustphase/details}} All observed phase angles lie on the Halley branch, $\Phi_{\rm H}(\alpha)=10^{-0.01807\alpha+0.000177\alpha^2}$, with $\alpha$ in degrees; the composite joins the Marcus component at $42^\circ$.

Following \citet{AHearn1984}, we compute
\begin{equation}
\af(\alpha)=\frac{4r_h^2\Delta^2}{\rho}
10^{-0.4(m_{\rho,\lambda}-m_{\odot,\lambda})},
\end{equation}
where $A$ is the effective scattering coefficient, $f$ the aperture filling factor, and $\rho$ the projected aperture radius. We express $r_h$ numerically in au, $\Delta$ and $\rho$ in cm, and the apparent solar magnitude $m_{\odot,\lambda}$ at 1 au in the corresponding photometric system. The zero-phase value is $A(0^{\circ})f\rho=\af(\alpha)/\Phi(\alpha)$. Appendix~\ref{app:phot} gives the passband-specific solar calibration. We report effective $\af$ because the aperture includes unresolved nucleus light and may include gas emission.

\section{Results}
\subsection{Archival recovery and activity constraints}\label{sec:archive}
Four CFH12K $R$-band exposures on 2000 September 23 covered the predicted position of N2 for a total of 1,560~s. In the stack aligned on the predicted motion, the aperture flux at that position was $0.71\sigma$ below the median blank-aperture flux. A brighter feature inside the orbit uncertainty ellipse remained fixed relative to the field stars on September 24. We find no secure CFHT recovery and cannot determine whether N2 was active in 2000.

The ten 150~s DECam $i$-band exposures on 2021 July 4 span UT 09:22:32--09:51:51 and were obtained for the Local Volume Complete Cluster Survey (LoVoCCS) (program 2019A-0308; principal investigator I. Dell'Antonio; \citealt{Fu2022LoVoCCS}). The headers identify I. Dell'Antonio and S. Fu as observers. At airmass 1.11, the per-frame median stellar FWHMs range from $1.03''$ to $1.41''$, with an overall median of $1.20''$. The clear-sky flags are false globally in all frames and locally in nine. We use the pipeline magnitude calibration for the N2-aligned stack (Appendix~\ref{app:decam}).

At true anomaly $\nu=+15.78^\circ$ we recovered N2 at the predicted position with $i=22.90\pm0.12$ mag and S/N $=9.33$. The $5\sigma$ sensitivity at blank positions, $i\simeq23.58$, applies only to a point source. A sky plane changes the measured flux by 12.4\% (0.13 mag) relative to a constant background. The Horizons total-magnitude prediction ($T\simeq22.39$) does not represent an inactive nucleus in $i$.

\begin{figure}[!htbp]
\centering
\includegraphics[trim=0bp 13bp 0bp 4bp,clip,width=\textwidth]{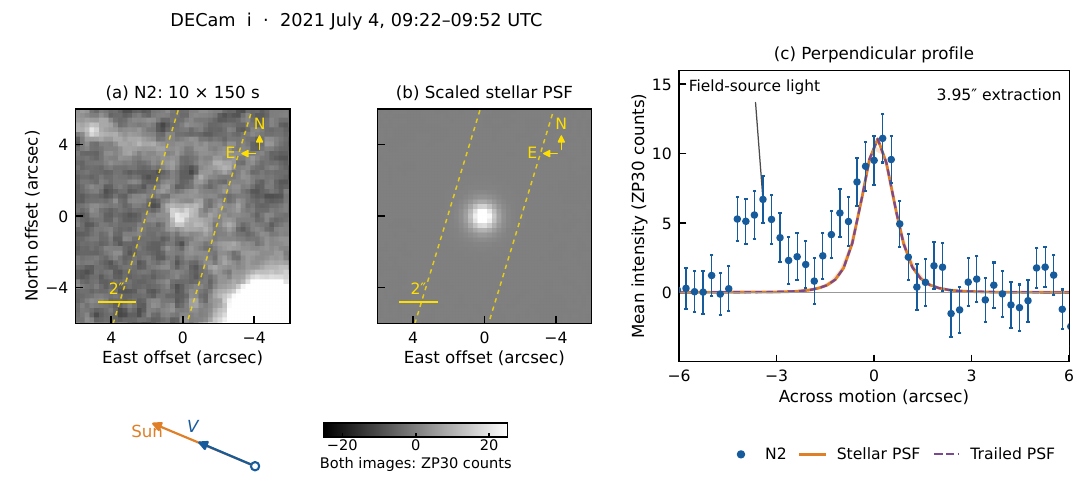}
\caption{DECam $i$-band comparison on 2021 July 4 ($10\times150$~s): (a) N2-aligned stack with N2 near the origin; (b) empirical PSF from 63 brighter, unsaturated field-star cutouts, scaled to N2's core; and (c) profiles perpendicular to the motion. The median stellar FWHM is $1.20''$, with individual-frame medians of $1.03''$--$1.41''$. Both images are north up and east left and share a linear count scale and grayscale stretch for clarity. Gaussian smoothing ($\sigma=0.18''$) is used only for display; profiles use unsmoothed pixels. Yellow dashed lines enclose the $3.95''$ strip along the motion. The positive heliocentric velocity (blue $V$) and sunward (orange) vectors have nearly identical position angles; their lengths are arbitrary. Orange and purple curves show the empirical PSF without and with the predicted $1.05''$ trail, respectively, with a fitted $0.11''$ core offset. Orange shading spans the 16th--84th percentiles of the scaled stellar profiles. Error bars show the empirical blank-profile scatter and exclude PSF, centroid, and calibration uncertainties. The feature near $-3.42''$ is associated with stationary field light.}\label{fig:decam}
\end{figure}

The predicted $1.05''$ trail per exposure is comparable to the $1.20''$ stellar FWHM. We therefore compare the perpendicular profile with 63 unsaturated stellar cutouts from the same exposures (Figure~\ref{fig:decam}; Appendix~\ref{app:decam}).

The mean wing excess at $1.5''$--$3''$ is $1.25\sigma$ using the empirical covariance, or $1.90\sigma$ on the negative side. Common-scale contours and the sidereal stack identify the feature at $-3.42''$ as stationary field light. A weak coma or narrow tail could remain unresolved. The structured sky and approximate calibration limit the photometry, so we exclude this epoch from the brightness comparison. Its unresolved profile does not establish an inactive-nucleus baseline.

We found no other convincing detections in the DECam images examined.

\subsection{Optical morphology and colors}
The June and July Simonyi images show dust before perihelion. After perihelion the Keck $g$ and $R$ profiles are broader than the stellar profiles (Figure~\ref{fig:psf}). The comet exceeds the empirical background by at least $3\sigma$ through $2.1''$--$2.4''$ in $g$ and $1.5''$--$1.8''$ in $R$. These radii depend on the sky subtraction, and changes in the stellar PSF limit the intrinsic-width measurement (Appendix~\ref{app:keck}).

Within a common projected aperture radius of 1,015.11 km ($\simeq0.69''$), we measure Keck colors of $g-r=0.50\pm0.05$, $r-i=0.06\pm0.06$, and $g-i=0.56\pm0.06$ mag. The transformed magnitude is $r=23.31\pm0.04$ mag (Equation~\ref{eq:rtransform}). Relative to the adopted solar colors, the two correlated colors are consistent with flat reflectance ($\chi^2=1.60$ for two degrees of freedom). The $g$-to-$i$ reflectance gradient, normalized by the mean of the endpoint reflectances, is $0.17^{+2.00}_{-1.98}$\% per 100 nm.

\begin{figure}[!htbp]
\noindent\begin{minipage}[c]{0.438\linewidth}\centering
\includegraphics[width=\linewidth]{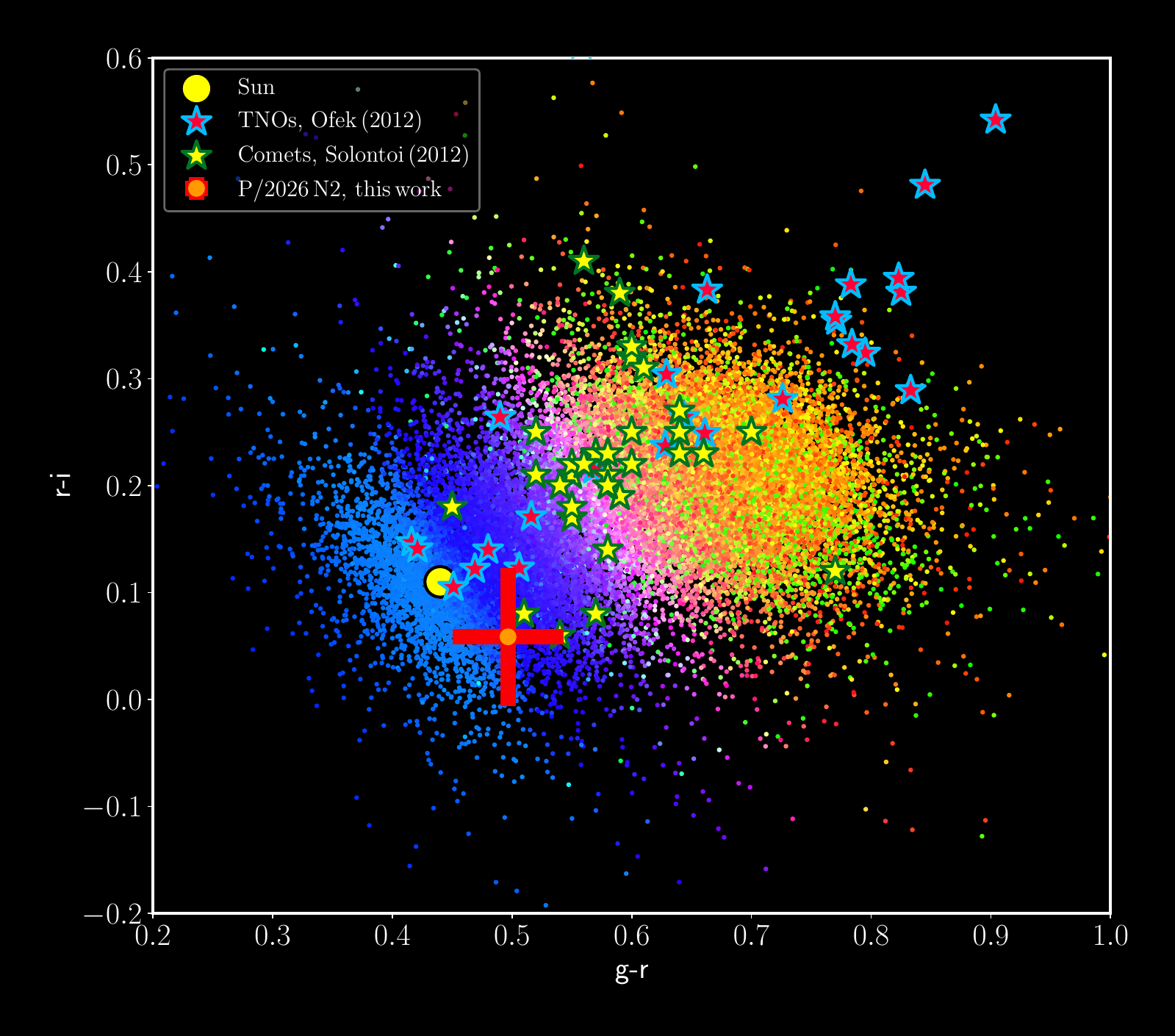}
\end{minipage}\hfill\begin{minipage}[c]{0.552\linewidth}\centering
\includegraphics[width=\linewidth]{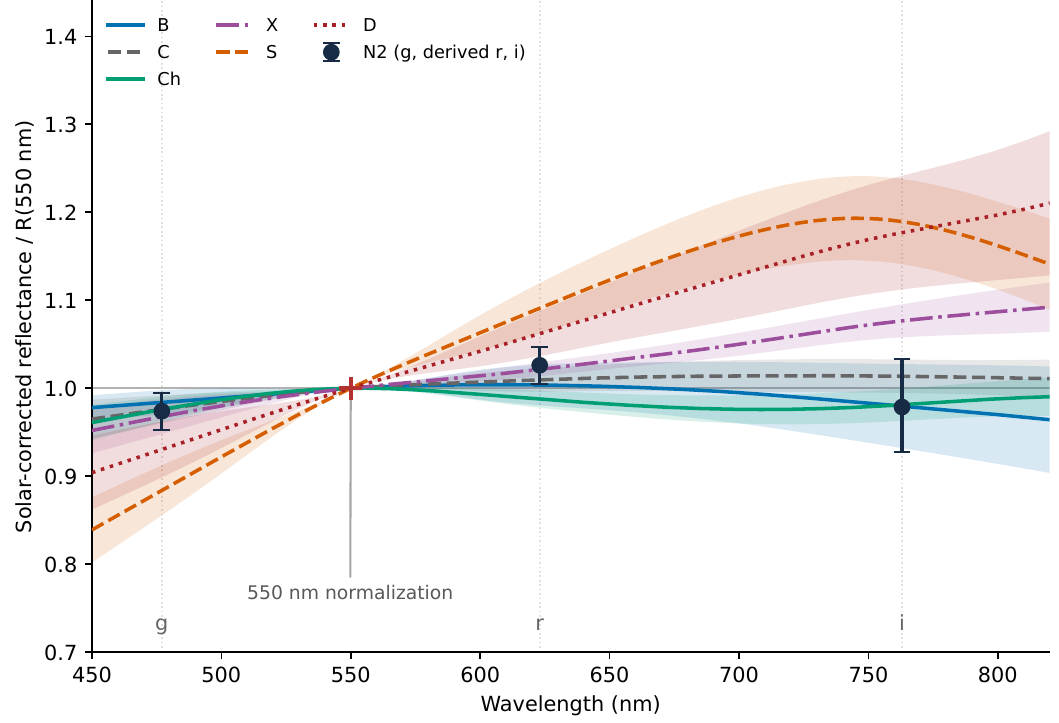}
\end{minipage}
\caption{\textit{Left:} The orange point and red error bars show the $g-r$ and $r-i$ colors of P/2026 N2; the yellow circle marks the Sun. Cyan-outlined stars show the trans-Neptunian comparison sample from \citet{Ofek2012}, and green-outlined stars show comets from \citet{Solontoi2012}. Small colored points show C-type (blue), S-type (red-orange), and V-type (green) comparison asteroids \citep[][]{Ivezic2001}. The Keck $r$ magnitude is transformed from Cousins $R_C$ and $g$, with the resulting covariance retained in the color uncertainties. \textit{Right:} Keck solar-corrected spectrophotometry of P/2026 N2 from UT 2026 October 7, within an aperture radius of 1,015.11 km ($\simeq0.6914''$). Points show $g$, derived $r$, and $i$ reflectances at nominal wavelengths 477, 623, and 763 nm, normalized at 550 nm. The cross marks the normalization rather than an additional measurement. Lines show selected Bus--DeMeo class means \citep{DeMeo2009}; translucent bands show their published pointwise $\pm1\sigma$ dispersion within each class. Both the mean and dispersion are divided by the mean reflectance at 550 nm. For display, cubic splines interpolate the original means and dispersions, with the dispersion constrained to zero value and slope at 550 nm. Fits use the original sampled templates and exclude the dispersion bands. Measurement error bars are 68\% photometric intervals propagated through the $R_C$-to-$r$ transformation and normalization with their covariance; solar-color, differential-seeing, and transformation systematics are excluded.}\label{fig:photometry}\label{fig:spectrum}
\end{figure}

The LSST July 9 measurements use $0.96''$ radii, corresponding to 1,008.10 km in $r$ and 1,008.04 km in $i$, and give $r-i=0.10\pm0.08$ mag. The Keck-minus-LSST color difference is $-0.04\pm0.10$ mag before any comet-spectrum passband transformation. Applying the respective solar corrections gives $i/r$ reflectance ratios of $0.95\pm0.06$ for Keck and $0.99\pm0.07$ for LSST. Neither comparison shows a significant color change (Figure~\ref{fig:photometry}).

The B-type Bus--DeMeo mean gives the smallest reduced $\chi^2$ (0.30), followed by C (0.38), Ch (0.47), and Cgh (0.64); X is also compatible (1.73), whereas D and S give 8.35 and 15.33. The class envelopes overlap (Figure~\ref{fig:spectrum}), and $gri$ samples only part of the 0.45--2.45 $\mu$m taxonomy \citep{DeMeo2009}. The combined coma and nucleus light thus resembles the nearly neutral B/C-complex templates. Appendix~\ref{app:phot} describes the sensitivity to the solar colors and passband transformation.

\subsection{Brightness around perihelion}\label{sec:brightness}
We derive effective $A(0^{\circ})f\rho=3.54\pm0.11$ cm in $g$, $3.56\pm0.14$ cm in native $R_C$, $3.46\pm0.17$ cm in $i$, and $3.66\pm0.13$ cm in derived SDSS $r$ at Keck. The Simonyi measurements give $3.32\pm0.16$ and $3.58\pm0.11$ cm in $r$ on July 9 and 14, respectively, and $3.03\pm0.26$, $3.05\pm0.17$, and $3.25\pm0.16$ cm in $i$ on June 27, July 6, and July 9. Within the $\sim$1,000 km apertures, the pre- and post-perihelion estimates are comparable under the adopted phase correction.

\begin{figure}[!htbp]
\centering\includegraphics[trim=0bp 18bp 0bp 16bp,clip,width=0.90\textwidth]{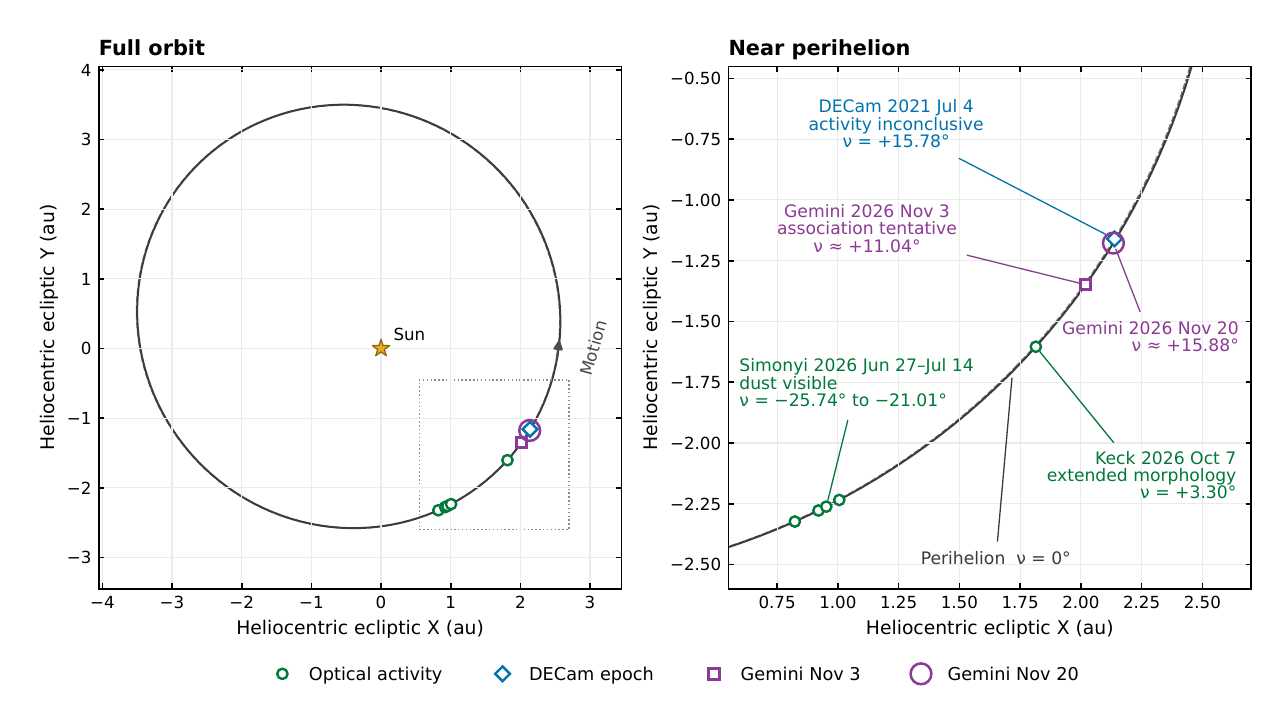}
\caption{Locations of the archival DECam and 2026 observations projected onto the heliocentric ecliptic plane using the JPL Horizons orbit. Dashed and solid curves show osculating orbits at 2021 July 4 and 2026 October 7, respectively. The right panel enlarges the near-perihelion region. Green circles indicate detected or reported activity in the Rubin and Keck data. The blue open diamond marks the DECam observation on 2021 July 4, for which the activity state is inconclusive. In both panels, the open purple square and circle mark the Gemini North pointings on November 3 and 20, respectively. We propagated the October 7 osculating orbit to the mean $J$ epoch of each night ($\nu\simeq+11.04^\circ$ and $+15.88^\circ$). The November 20 $Y$ position coincides with the $J$ position at this scale. The Gemini symbols mark the pointing geometry, irrespective of whether the compact features are N2.}\label{fig:activity}
\end{figure}

From July 9 to October 7, $i$ faded from $21.98\pm0.05$ to $23.25\pm0.05$ mag. After correcting for geometry and phase, the Keck aperture was $0.07\pm0.07$ mag brighter ($H_{\rho,i}=19.04\pm0.05$ and $18.97\pm0.05$ mag, respectively; Table~\ref{tab:photo}). The July 9 and Keck $H_{\rho,r}$ values of $19.14\pm0.05$ and $19.03\pm0.04$ mag differ by $0.11\pm0.07$ mag. The inverse-variance-weighted Simonyi means are $H_{\rho,r}=19.09\pm0.03$ and $H_{\rho,i}=19.07\pm0.04$ mag. The June and July Simonyi observations span true anomalies $-25.74^\circ$ to $-21.01^\circ$ (90--74 days before perihelion); Keck observed at $+3.30^\circ$, 11.5 days after perihelion (Figure~\ref{fig:activity}). The November 3 and 20 Gemini pointings were at $\nu\simeq+11.04^\circ$ and $+15.88^\circ$, respectively.

At a projected radius of 2,000 km, we compare native LSST $r$ with Keck $r$ derived from $g$ and $R_C$ measured at that radius (Table~\ref{tab:largeap}). Both data sets fully cover the aperture and the 2,300--2,900 km sky annulus. With this sky annulus, the corrected October Keck magnitude minus the inverse-variance-weighted LSST mean for July 9 and 14 is $-0.10\pm0.09$ mag, consistent with no change within the formal errors.

\begin{table}[!htbp]
\centering
\caption{Brightness and effective dust measurements within a 2,000 km aperture radius.}\label{tab:largeap}
\small\setlength{\tabcolsep}{8pt}
\begin{tabular}{rllrr}
\toprule
$\rho$ (km) & UT date (2026) & Facility & $H_{\rho,r}$ (mag) & $A(0^{\circ})f\rho$ (cm)\\
\midrule
2,000 & Jul 9  & LSST & $18.85\pm0.11$ & $2.19\pm0.23$\\
2,000 & Jul 14 & LSST & $18.67\pm0.07$ & $2.59\pm0.16$\\
2,000 & Oct 7  & Keck  & $18.61\pm0.07$ & $2.74\pm0.18$\\
\bottomrule
\end{tabular}
\par\vspace{2pt}\begin{minipage}{0.98\textwidth}\footnotesize
Measurements use a 2,000 km aperture radius and a fully sampled 2,300--2,900 km sky annulus. The July LSST measurements use native $r$; Keck $r$ is derived from $g$ and $R_C$ with Equation~\ref{eq:rtransform} and the SDSS solar reference $M_{\odot,r}=4.65$ AB mag. Both quantities use the Halley--Marcus phase correction and include nucleus light. Errors include formal photometric terms but exclude phase-law, solar-reference, passband, PSF, and background systematics.
\end{minipage}
\end{table}

Seeing differences can change the fraction of light inside the matched apertures, and the stellar $R_C$-to-$r$ transformation has an unquantified uncertainty for the comet spectrum. Appendix~\ref{app:phot} quantifies the sensitivity to centering.

On November 3 the compact feature selected in the $J$ images lay $0.56''$ from the predicted position, inside the combined Horizons and WCS $3\sigma$ ellipse (Figure~\ref{fig:gnirs}). At this position we measured $61.4\pm17.5$ electrons s$^{-1}$ (pointwise S/N~3.5), or $J_{\rm MKO,Vega}=22.50^{+0.36}_{-0.27}$ mag, with a separate calibration floor of $0.06$ mag. At the exact November 3 prediction the fit gave $24.1\pm17.1$ electrons s$^{-1}$; the measured flux plus three empirical standard errors gives $J_{\rm MKO,Vega}>22.28$ mag for compact emission at that position.

On November 20 the $Y$ peak inside the combined $3\sigma$ ellipse lay $0.45''$ west and $0.30''$ south of the Horizons prediction. At this fixed offset, fits to individual unsmoothed exposures gave $J_{\rm MKO,Vega}=23.15^{+0.44}_{-0.31}$ and $Y_{\rm Vega}=23.43^{+0.27}_{-0.22}$ mag, with pointwise S/N of 3.02 and 4.55, respectively (Table~\ref{tab:photo}). The magnitude errors are statistical; the separate calibration floors are 0.026 mag in $J$ and 0.040 mag in $Y$. At the exact predicted position the same fits gave S/N of 0.46 and 3.07. The $J$ and $Y$ peaks were $0.21''$ apart, and both halves of each sequence gave positive flux at the selected offset. We regard the compact feature as a tentative counterpart: these significances are conditional on position and do not account for selecting the peak in $Y$.

From the November 20 native-frame fluxes we obtain an apparent Vega color of $Y-J=+0.29^{+0.41}_{-0.48}$ mag. The errors give the 68.27\% interval for the flux ratio (Appendix~\ref{app:gnirs}); independent zero-point errors would contribute a further 0.048 mag of calibration uncertainty. The two sequences were taken about half an hour apart. Selecting the position in $Y$ can bias its flux high, and the GNIRS-to-Hillenbrand $Y$ color term for N2 is unknown.

To compare the two $J$ nights, we assumed constant intrinsic activity over the 17 days and applied the Halley--Marcus dust phase function to the compact emission. At the November 20 distances and phase angle, the November 3 image-selected feature would have $J_{\rm MKO,Vega}=22.66^{+0.36}_{-0.27}$ mag (statistical), 0.16 mag fainter than observed on November 3. We registered the two images on the predicted positions, combined them, and measured the coadd at the common November 20 $Y$ offset (Figure~\ref{fig:gnirscoadd}; Appendix~\ref{app:gnirs}). In a $0.60''$-radius aperture the coadd gives $33.5\pm8.7$ electrons s$^{-1}$ on the November 20 scale (pointwise S/N~3.86); at the exact prediction its S/N is 2.97. At the $Y$ offset the individual nights give S/N~1.31 and 4.27, respectively. Adding November 3 does not strengthen the November 20 result.

In matched apertures, November 20 $Y$ minus the two-night $J$ coadd at November 20 geometry gives $+0.26^{+0.39}_{-0.40}$ mag (68.27\% statistical interval). This check assumes comparable fractions of compact light inside the apertures. Both color intervals include zero. These compact-source colors do not constrain composition or intrinsic variability, and the apertures do not measure total coma flux.

\subsection{Orbit and forward evolution}\label{sec:orbit}
We adopt the JPL Horizons solution dated 2026 August 25, 07:42:02, retrieved on 2026 October 14 \citep{Giorgini1996}.\footnote{\url{https://ssd.jpl.nasa.gov/horizons/}} Table~\ref{tab:orbit} lists its heliocentric ecliptic J2000 osculating elements and formal $1\sigma$ uncertainties at Julian date (JD) 2460616.5 in barycentric dynamical time (TDB) (2024 November 2). The orbit has $a=3.07$ au, $e=0.21$, and $i=2.25^\circ$, with approximate Jupiter Tisserand parameter $T_J=3.19$. We tested orbital stability by integrating 5,000 clones forward for 50~Myr (Appendix~\ref{app:integrations}).

\begin{table}[!htbp]
\centering
\caption{JPL Horizons orbital elements of P/2026 N2 and formal uncertainties.}\label{tab:orbit}
\small\setlength{\tabcolsep}{12pt}
\begingroup\let\tabular\savetabular
\begin{tabular}{ll}
\toprule
Parameter & Value ($1\sigma$)\\
\midrule
Semimajor axis $a$ (au) & $3.0725\pm(1.35\times10^{-5})$ \\
Eccentricity $e$ & $0.2123\pm(2.72\times10^{-5})$ \\
Inclination $i$ (deg) & $2.2497\pm(6.09\times10^{-5})$ \\
Perihelion distance $q$ (au) & $2.4203\pm(9.41\times10^{-5})$ \\
Longitude of ascending node $\Omega$ (deg) & $96.382\pm(8.38\times10^{-4})$ \\
Argument of perihelion $\omega$ (deg) & $219.03\pm(8.18\times10^{-3})$ \\
Mean anomaly $M$ (deg) & $233.13\pm(5.20\times10^{-3})$ \\
Orbital period $P$ (yr) & $5.3857\pm(3.55\times10^{-5})$ \\
Jupiter Tisserand parameter $T_J$ & $3.1944\pm(4.92\times10^{-6})$ \\
\bottomrule
\end{tabular}\par\endgroup
\par\vspace{2pt}\begin{minipage}{0.98\textwidth}\footnotesize
Heliocentric ecliptic J2000 osculating elements at JD 2460616.5 TDB (2024 November 2), from the JPL Horizons solution dated 2026 August 25, 07:42:02, retrieved on 2026 October 14 \citep{Giorgini1996}. Formal $1\sigma$ uncertainties are from the JPL Small-Body Database for the same solution and epoch (\url{https://ssd.jpl.nasa.gov/tools/sbdb_lookup.html}). The period and its uncertainty use 365.25 days per year. We calculate $T_J=a_J/a+2\cos i\sqrt{(a/a_J)(1-e^2)}$ with $a_J=5.2044$ au; its uncertainty is propagated from the full JPL orbital covariance with $a_J$ held fixed. The integrations use the full-precision solution.
\end{minipage}
\end{table}

\begin{figure}[!htbp]
\centering\includegraphics[trim=0bp 14bp 0bp 13bp,clip,width=0.87\textwidth,height=0.78\textheight,keepaspectratio]{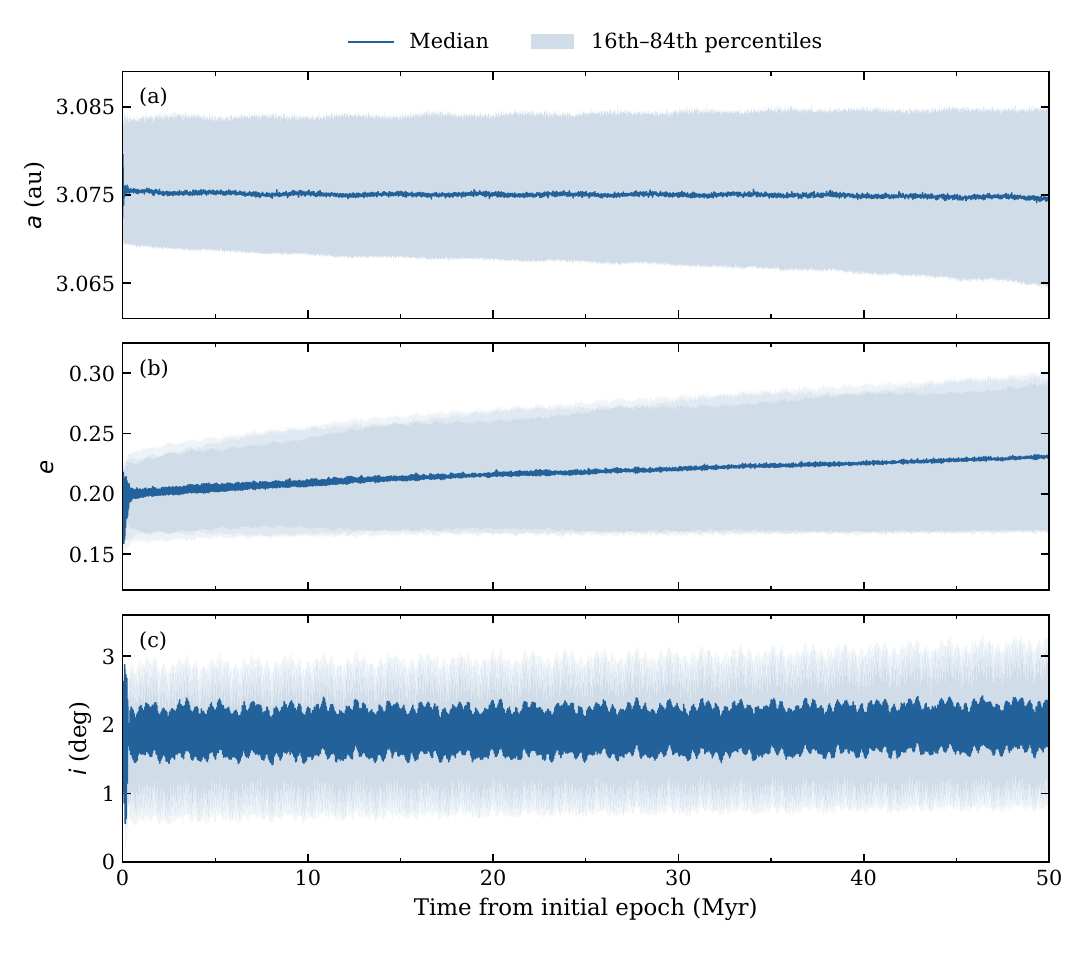}
\caption{Forward evolution of all 5,000 clones of N2 over 50 Myr. From top to bottom: heliocentric osculating semimajor axis, eccentricity, and inclination relative to ecliptic J2000. Curves show the unsmoothed median; shading spans the pointwise 16th--84th percentiles. Time is measured from JD 2460616.5 TDB, with outputs every 10 kyr.}\label{fig:dynamics}
\end{figure}

At each output epoch, the 16th and 84th percentiles of semimajor axis for all 5,000 clones fall between 3.064 and 3.086 au (Figure~\ref{fig:dynamics}). At 50~Myr the median eccentricity and inclination are 0.229 and $2.34^\circ$. At 25 and 50~Myr, respectively, 4,989 (99.78\%) and 4,770 (95.40\%) of the original clones are bound ($a>0$, $e<1$). The narrow core indicates long-term stability of N2's present orbital neighborhood under the adopted force model, consistent with prolonged past residence in the outer Main Belt.

\subsection{Projected dust geometry}\label{sec:fp}
\begin{figure}[!htbp]
\centering
\includegraphics[trim=0bp 8bp 0bp 4bp,clip,width=\textwidth]{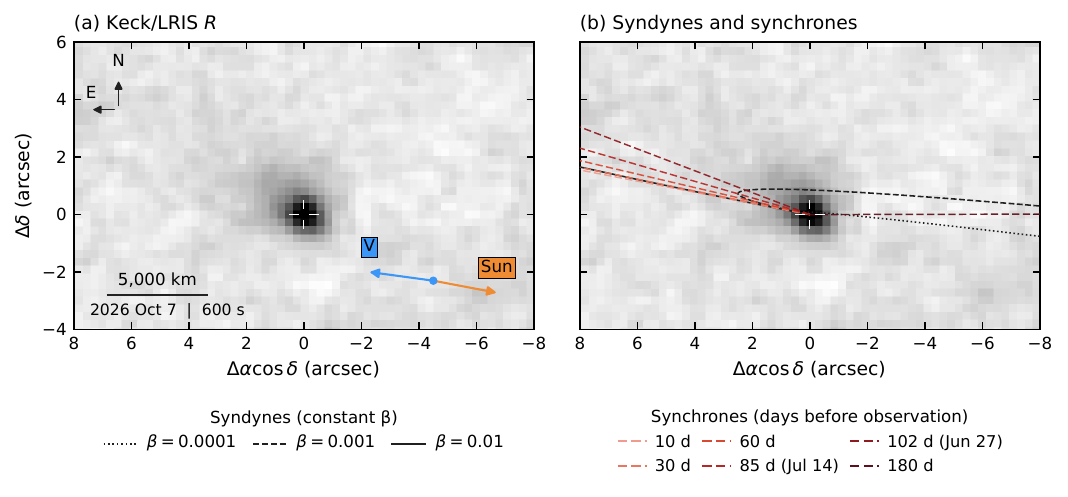}
\caption{Keck/LRIS $R$ image of \obj\ on UT 2026 October 7, combining two 300 s exposures: (a) sky-subtracted image and (b) the same image with zero-ejection-speed syndynes (black, constant $\beta$) and synchrones (red, common release time). Syndynes use $\beta=10^{-4}$, $10^{-3}$, and $10^{-2}$ over release ages of 0--365 days. Synchrone labels give ages of 10--180 days, with $0\leq\beta\leq0.1$; 85 and 102 days approximately correspond to July 14 and June 27. Both panels use the same linear stretch and Gaussian display smoothing of $\sigma=0.18''$; grayscale makes the overlays more visible. The cross marks the coma photocenter. North, east, and a 5,000 km scale are shown. Orange Sun and blue V arrows indicate the sunward and positive heliocentric-velocity directions, with arbitrary lengths and a common blue circular base. The curves show projected trajectories; no dust-production history is fitted.}\label{fig:dust}
\end{figure}

We compare the Keck tail with zero-ejection-speed syndynes and synchrones \citep{Finson1968,Hsieh2015a}. We propagate grains under $(1-\beta)$ times solar gravity, where $\beta$ is the radiation-pressure-to-gravity ratio, using the August 25 JPL Horizons solution at the exposure-weighted epoch (Figure~\ref{fig:dust}). Viewed $1.10^\circ$ above the orbital plane, the 85- and 102-day synchrones separate by just $0.16''$ at $2''$ and $0.41''$ at $5''$ from the photocenter. Neither separation exceeds the approximately $1''$ core width, and no ridge in the outer tail distinguishes the two release dates. Grains of different sizes from a single release can span several syndynes. Comparable models inferred emission lasting at least three months for 313P/Gibbs \citep{Hsieh2015a} and 6--9 months or longer for both components of P/2016 J1 \citep{Moreno2017}. For quasi-Hilda comet 282P, activity persisted for more than 15 months and recurred at successive perihelia \citep{Chandler2022_282P}.

\subsection{Sunward coma width and dust scales}
\label{sec:dustscales}
\begin{figure}[!htbp]
\centering
\includegraphics[trim=0bp 9bp 0bp 8bp,clip,width=0.94\textwidth]{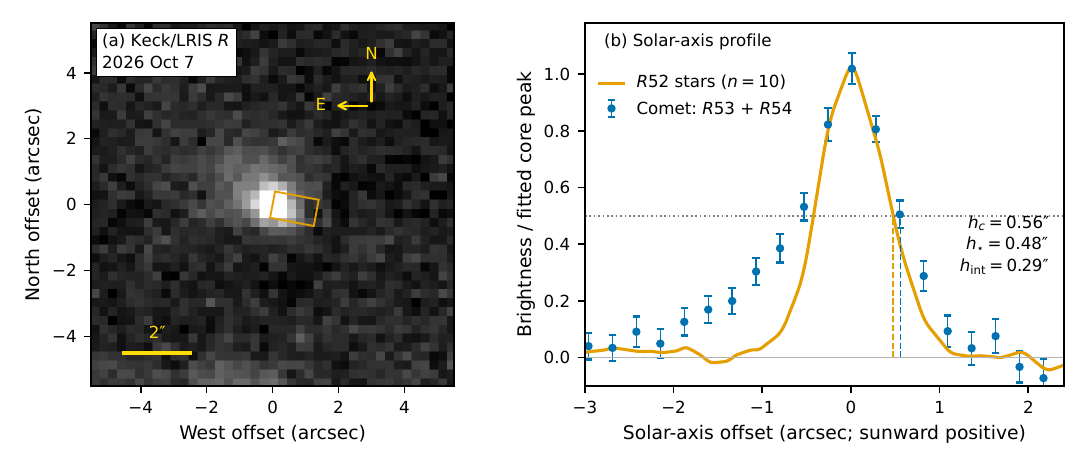}
\caption{Sunward profiles in the Keck/LRIS $R$ data from 2026 October 7. Left: unsmoothed summed image in grayscale for clarity; the orange rectangle marks the sunward $0$--$1.35''$ segment of the $0.81''$-wide extraction strip. Right: profiles of N2 and ten stars in sidereal reference frame 52, normalized to their fitted core peaks, with positive offsets toward the Sun. Quadrature subtraction of their half-maximum widths gives the approximately 420-km projected core scale used in Equation~\ref{eq:sunward_velocity}; it does not locate a physical stopping distance. Profiles use unsmoothed pixels. Error bars show blank-profile scatter divided by the fitted core peak and exclude normalization and PSF uncertainty.}\label{fig:sunward}
\end{figure}

In Keck $R$, the sunward half-width at half maximum is $h_c=0.56''$ for N2 and $h_\star=0.48''$ for the reference stars (Figure~\ref{fig:sunward}; Appendix~\ref{app:keck}). Gaussian quadrature subtraction gives $h_{\rm int}=(h_c^2-h_\star^2)^{1/2}=0.29''$, or $D_{\rm core}\simeq420$ km. Resampling stars and source noise gives a 16th--84th percentile range of 290--520 km; strips $0.27''$--$1.35''$ wide give $h_{\rm int}=0.26''$--$0.37''$. These intervals exclude temporal PSF changes, tracking, and non-Gaussian profile shapes.

The measured core width does not establish the dust turnaround point. We express the assumed sunward stopping distance as $D_t=\eta D_{\rm core}$ \citep{Hsieh2021QN173,Bolin20253I} and use the radiation-pressure relation \citep{Jewitt1987} to obtain the terminal sunward speed
\begin{equation}
v_e=\frac{\sqrt{2\beta g_\odot\eta D_{\rm core}}}{r_h}
\simeq29.4\sqrt{\beta\eta}\ {\rm m\,s^{-1}},
\label{eq:sunward_velocity}
\end{equation}
where $g_\odot=0.006$ m s$^{-2}$ is the solar gravitational acceleration at 1 au, $r_h=2.423$ in au, and $D_{\rm core}$ is in meters. For $\eta=1$ and $\beta\simeq(a_d/1\,\mu{\rm m})^{-1}$, grain radii $a_d=1~\mu$m and 1~mm give speeds of 29 and 0.93 m s$^{-1}$, respectively. The image does not constrain $\eta$, so these speeds are model estimates rather than observational limits. If $D_{\rm core}$ is the projected stopping distance along the Sun--comet axis, then $\eta=1/\sin\alpha$ and both model speeds increase by a factor of 1.57 at $\alpha=23.89^\circ$.

We estimate a dust mass-loss rate from the measured $R_C$ flux assuming an optically thin, steady, isotropic coma of identical grains of radius $a_d$, density $\rho_d$, and constant speed $v_e$ \citep{Fink2012}. A circular aperture of radius $\rho$ samples a cylinder through the coma, giving a mean residence time $\pi\rho/(2v_e)$. In our geometric-albedo convention $A(0^\circ)=4p_V$. For a number production rate $Q_N$, Appendix~\ref{app:mass} gives
\begin{equation}
\dot M=\frac{4\pi}{3}\rho_da_d^3Q_N
=\frac{2\rho_da_dv_e}{3p_V}\,A(0^\circ)f\rho,
\label{eq:dust_rate}
\end{equation}
consistent with \citet{Kiss2015SidingSpring}.

To evaluate Equation~\ref{eq:dust_rate} for the Keck $R_C$ measurement, we adopt $\rho_d=750$ kg m$^{-3}$ and $p_V=0.10$ \citep{Bolin20253I}; neither is measured for N2. Applying $p_V$ to $R_C$ assumes approximately gray $V$--$R_C$ scattering. If dust supplies all of the native $R_C$ signal, $A(0^\circ)f\rho=3.56\pm0.14$ cm gives $\dot M\simeq0.01$ kg s$^{-1}$ for $1~\mu$m grains and $0.17$ kg s$^{-1}$ for 1~mm grains at the respective $\eta=1$ speeds. At fixed size and speed, the rates scale as $\rho_d/p_V$. When the speed is calculated with Equation~\ref{eq:sunward_velocity}, they also scale as $\eta^{1/2}$.

The conversion assumes approximately constant production over the grains' residence time. The photometric error does not account for the unknown nucleus contribution, aperture losses, phase-law uncertainty, or departures from isotropic radial flow caused by anisotropic emission or radiation pressure.

\subsection{Impulsive dust retention}\label{sec:retention}
We assess whether dust from a single release could account for the matched-aperture photometry. Following \citet{Luu2026K1}, we model an impulsive release on June 27 with zero ejection speed relative to N2. To lie within the projected 2,000~km aperture at the observing epochs, grains must have radii of at least about $30~\mu$m on July 9, $47~\mu$m on July 14, and 1.7~mm on October 7 (Appendix~\ref{app:retain}, Equations~\eqref{eq:impulse_motion}--\eqref{eq:impulse_cutoff}).

In the 2,000~km aperture, the corrected Keck $r$ magnitude in October minus the weighted LSST $r$ magnitude in July is $-0.10\pm0.09$ mag (Table~\ref{tab:largeap}). The central value is about 10\% brighter in October. As a fading tolerance we adopt $+0.08$ mag, two formal uncertainties above the measured difference. If dust supplies all the aperture light and its scattering properties stay fixed, at least 93\% of the July scattering area must remain to satisfy this tolerance (Appendix~\ref{app:retain}, Equation~\eqref{eq:impulse_fraction}). For a power law $dN/da_d\propto a_d^{-s}$ extending down to $1~\mu$m, this requires maximum radii of at least 2.4 cm for $s=2$ or 25 cm for $s=2.5$. A broad $s=3.5$ distribution would fade by about 2.0 mag; the same slope can reproduce the photometry if grains smaller than about 1.6 mm were absent for a 1 cm upper cutoff. Small grains could have been present at release if they contributed little of the July aperture light.

Models of the 2014 tail of 313P/Gibbs indicate that grains with radii of 50--100~$\mu$m dominated its optical appearance \citep{Jewitt2015_313P}; the resolved 2011 tail of 288P was fit with grain radii of 10--300~$\mu$m \citep{Agarwal2016_288P}. A June 27 release dominated in July by grains of these sizes would fade more than N2's photometry allows. Millimeter grains are plausible: in 2016, 288P briefly released millimeter-sized grains before grains of about $10~\mu$m dominated its visible tail \citep{Agarwal2017_288Pbinary}. For the 2013 tail of 133P/Elst-Pizarro, \citet{Jewitt2014EPPP} derived effective mean radii of 60--90~$\mu$m or 2--3~mm for continuous emission over 36 or 150 days, respectively. Figure~\ref{fig:activeae} marks the orbits of these three comets.

\section{Discussion and Conclusions}
The Simonyi and Keck images resolve dust on both sides of perihelion over 102 days. Sustained or recurrent dust production can explain the July and October brightnesses, which agree within the formal errors in matched apertures (Section~\ref{sec:brightness}); $A(0^\circ)f\rho$ traces the same flux. A single June 27 release at zero ejection speed can also match the brightness if grains of radius $\gtrsim1.7$~mm already supplied at least about 93\% of the July scattering area, assuming that dust supplied all the aperture light and using the fading tolerance in Section~\ref{sec:retention}. The November Gemini measurements probe a tentative compact counterpart; October 7 remains the last epoch in which we resolve dust.

Impact ejecta can fade rapidly or persist: (596)~Scheila faded by about 30\% in eight days \citep{Jewitt2011Scheila}, whereas a single collision reproduced the debris of P/2010~A2 \citep{Kleyna2013A2} and its large grains remained visible a full orbit later \citep{Jewitt2013}. (6478)~Gault was active over six years, including near aphelion \citep{Chandler2019Gault}. Its discrete low-speed releases \citep{Ye2019Gault,Hui2019Gault,Kleyna2019Gault} and approximately 2.5~h quiescent rotation period \citep{LuuGaultrotation2021,Purdum2021} favor rotational shedding. The duration of N2's visible activity alone does not distinguish these mechanisms.

From the Keck $R_C$ photometry of N2, we infer dust mass-loss rates of approximately 0.01 and 0.17~kg~s$^{-1}$ for grain radii of $1~\mu$m and 1~mm, respectively (Equation~\ref{eq:dust_rate}). For 133P/Elst-Pizarro, \citet{Jewitt2014EPPP} inferred higher mean rates of 0.20 and 1.40~kg~s$^{-1}$ for continuous dust emission over 36 and 150 days, respectively. These rates depend on different grain populations: we assume one grain size per model for N2, whereas the 133P tail models fit a distribution of sizes over the emission interval.

Archival searches by citizen scientists have recovered previously unrecognized activity in DECam images \citep{Chandler2024}. Machine learning extends this search to survey streams \citep{Bolin2025AICometBook}: \texttt{Tails} detects activity in Zwicky Transient Facility (ZTF) data \citep{Duev2021Tails}, including twilight observations \citep{Bolin2025TS}, and \texttt{TailNet} with citizen-science inspection identified activity in 2016 UU121 and (410590) 2008 GB140 \citep{Sedaghat2024,Chandler2024GB140}. These methods could identify active objects in LSST twilight images. Our Keck follow-up added $g-r$ to the survey $r-i$ measurement, extending the multiband approach used for C/2022 E3 (ZTF) \citep{Bolin2024E3}.

\begin{figure}[!htbp]
\centering
\includegraphics[trim=0bp 14bp 0bp 8bp,clip,width=0.91\textwidth]{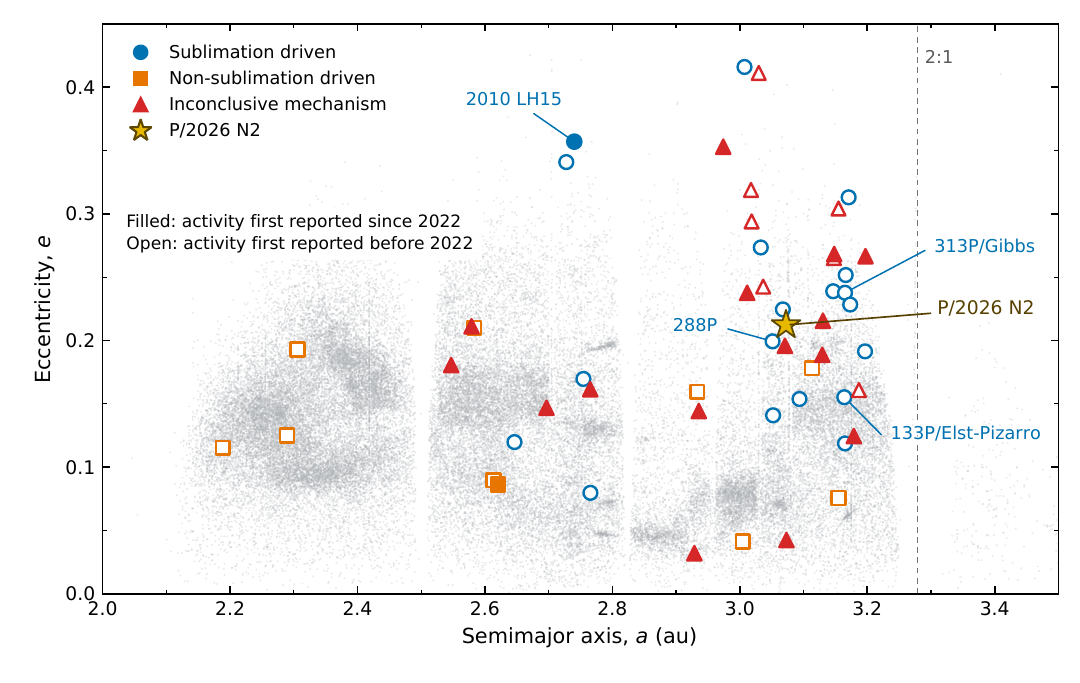}
\caption{Active objects in the Main Belt, following Figure~12 of \citet{Jewitt2022MBCscometsIII} and updated through 2026 October 12. Blue circles and orange squares denote sublimation and non-sublimation mechanisms favored in the literature, respectively; blue includes sublimation combined with rotation. Red triangles mark inconclusive cases. Filled symbols mark 17 objects whose activity was first reported since 2022, including archival discoveries; open symbols mark 34 earlier discoveries, with updated classifications. Each object or disrupted system appears once. The labeled 2010 LH15 is a recurrently active comparison. Labels also identify 133P/Elst-Pizarro, 288P, and 313P/Gibbs, whose dust sizes we compare in Section~\ref{sec:retention}. The gold star marks N2's orbit (Table~\ref{tab:orbit}). Comparison objects satisfy $2<a<3.5$ au, $q>1.666$ au, and $T_J>3.08$ and use osculating elements retrieved from the JPL Small-Body Database on 2026 October 12 at their catalog epochs. Gray points show AstDyS synthetic proper elements for asteroids numbered 1--60,000 \citep{Knezevic2000,Knezevic2003}; 56,701 lie within the displayed limits. The proper elements are from the June 2024 catalog, retrieved on 2026 October 12, without a discovery-bias correction. The dashed line marks the approximate 2:1 resonance with Jupiter. The accompanying source table lists object identities, orbital epochs, classification evidence, and references.}\label{fig:activeae}
\end{figure}

The stable core of the forward ensemble is consistent with prolonged Main Belt residence; neither the forward integration nor individual backward trajectories through chaotic encounters determine how long N2 has occupied this orbit \citep{Morbidelli2020}. Figure~\ref{fig:activeae} compares N2 with other active Main Belt objects, updating the compilation of \citet{Jewitt2022MBCscometsIII} with activity discoveries since 2022 \citep{Deen2023BJ22,Chandler2024,Chandler2024GB140,Sedaghat2024,Hsieh2026VZ8}. Recurrent activity supports sublimation in 2010 LH15 \citep{Chandler2024} and 456P/PANSTARRS \citep{Hsieh2024Recurrent456P}. N2's $a=3.07$ au and $e=0.21$ overlap this population, although similar orbits also host objects with other activity mechanisms. The background uses Asteroids Dynamic Site (AstDyS) proper elements; symbols for active objects use osculating elements.\footnote{\url{https://newton.spacedys.com/~astdys2/propsynth/all.syn}}

N2's nearly neutral B/C-like reflectance also resembles the B-type visible spectra of the Main Belt comets 133P/Elst-Pizarro and 176P/LINEAR \citep{Licandro2011MBCspectra}. For 133P, \citet{Hsieh2026_133P} detected water vapor at two epochs with the Near Infrared Spectrograph on the James Webb Space Telescope, providing a direct example of ice sublimation in a Main Belt comet with B-like reflectance. Our optical broadband colors of N2 combine coma and nucleus light.

The 2021 DECam images recover N2 but do not establish activity at that return. N2's outer Main Belt orbit, B/C-like reflectance, and comparable aperture brightnesses across perihelion together favor sublimation. We therefore classify N2 as a probable MBC. A decline away from perihelion followed by renewed activity at a future perihelion would favor fresh emission over lingering grains; a detection of water vapor would test the sublimation hypothesis directly.

\begin{acknowledgments}
We thank Matthew Belyakov for helping us secure Gemini observing time and the Gemini North staff for obtaining the GNIRS observations under Fast Turnaround program GN-2026B-FT-207. The International Gemini Observatory is a program of the U.S. National Science Foundation (NSF) through NOIRLab. The Association of Universities for Research in Astronomy operates the observatory under a cooperative agreement with the U.S. National Science Foundation on behalf of the Gemini partners: Argentina, Brazil, Canada, Chile, the Republic of Korea, and the United States.

This publication uses both pre-LSST and LSST data from NSF--DOE Vera C. Rubin Observatory, accessed as public alert-packet cutouts through the ALeRCE broker.

This material is based upon work supported in part by the National Science Foundation through Cooperative Agreements AST-1258333 and AST-2241526 and Cooperative Support Agreements AST-1202910 and 2211468 managed by the Association of Universities for Research in Astronomy (AURA), and the Department of Energy under Contract No. DE-AC02-76SF00515 with the SLAC National Accelerator Laboratory managed by Stanford University. Additional Rubin Observatory funding comes from private donations, grants to universities, and in-kind support from LSST-DA Institutional Members.

This research uses services or data provided by the Rubin Science Platform at NSF-DOE Vera C. Rubin Observatory, which is jointly funded by the U.S. National Science Foundation and the U.S. Department of Energy, Office of Science.

Some of the data presented herein were obtained at Keck Observatory, which is a private 501(c)3 non-profit organization operated as a scientific partnership among the California Institute of Technology, the University of California, and the National Aeronautics and Space Administration. The Observatory was made possible by the generous financial support of the W. M. Keck Foundation. The raw Keck Observatory data presented herein, once public, can be accessed from the Keck Observatory Archive (KOA), which is operated by the W. M. Keck Observatory and the NASA Exoplanet Science Institute (NExScI), under contract with the National Aeronautics and Space Administration.

The authors wish to recognize and acknowledge the very significant cultural role and reverence that the summit of Maunakea has always had within the Native Hawaiian community. We are most fortunate to have the opportunity to conduct observations from this mountain.

We acknowledge computational resources provided by the Bura supercomputer at the Center for Advanced Computing and Modelling, University of Rijeka, Croatia. We thank James Westover for assistance with setting up access to Bura.

We used OpenAI ChatGPT/Codex to assist with manuscript preparation and editing, development of analysis code, and figure preparation. The authors take responsibility for the final manuscript.
We thank the CFHT staff for the archival CFH12K observations and the Canadian Astronomy Data Centre for access to the images.
\end{acknowledgments}

\software{Data Reduction for Astronomy from Gemini Observatory North and South (DRAGONS) 4.2.2 \citep{Labrie2023DRAGONS}, LPipe \citep{Perley2019}, REBOUND 5.2.2 \citep{Rein2012REBOUND}, MERCURIUS \citep{Rein2019MERCURIUS}, IAS15 \citep{Rein2015IAS15}, Astropy \citep{Astropy2022}, NumPy \citep{NumPyHarris2020}, SciPy \citep{SciPyVirtanen2020}, Matplotlib \citep{MatplotlibHunter2007}, Photutils \citep{PhotutilsConcept}.}

\facilities{CFHT (CFH12K), Gemini:Gillett (GNIRS), Rubin:Simonyi (LSSTCam), Keck:I (LRIS), Blanco (DECam)}

\appendix
\renewcommand{\theHequation}{app.\Alph{section}.\arabic{equation}}
\section{Image reduction and photometric measurements}\label{app:reduction}
\subsection{DECam stacking and stellar profiles}\label{app:decam}
We subtract the sky, scale exposures to a common zeropoint using the archival pipeline calibration, and combine the N2-aligned frames with inverse-variance weights.

We calculate the trail in each sidereally tracked exposure as $\ell=t_{\rm exp}\sqrt{(\dot\alpha\cos\delta)^2+\dot\delta^2}/3600$, with $t_{\rm exp}$ in seconds and angular rates in arcsec h$^{-1}$. On July 4, the exposure-weighted eastward and northward rates are $24.12$ and $7.36$ arcsec h$^{-1}$, giving a total rate of $25.22$ arcsec h$^{-1}$ at position angle $73.04^\circ$ east of north. The trail in each 150~s exposure is therefore $1.05''$, or 4.00 pixels at $0.263''$ pixel$^{-1}$. Aligning the ten exposures on N2 retains the 150~s trail in the 1,500~s stack. Stationary sources shift by $11.28''$ relative to N2 between the first and last midpoints and can produce extended features in the stack.

To measure extension independently of trailing, we average the comet and same-exposure stellar images over a $3.95''$ strip along the motion and compare the resulting perpendicular profiles \citep{Hsieh2014}. We measure the sky at perpendicular offsets of $4.5''$--$6.5''$. The reference PSF combines 63 brighter, unsaturated stellar cutouts ($i\simeq16.8$--$17.9$ in the pipeline calibration; Figure~\ref{fig:decam}). We fit its amplitude and perpendicular centroid within $1.5''$ of N2's core and display both images with the same calibrated count scale and stretch. The fitted centroid shift is $0.11''$; adding the model to blank profiles and refitting gives a scatter of $0.07''$.

\subsection{Simonyi alert cutouts}\label{app:rubin}
We measured fluxes and propagated variances from the alert-packet science cutouts; Figure~\ref{fig:rubin} also shows the corresponding templates and difference images.

We use the UTC alert epochs, which agree with the MPEC times to within 0.51 s. For these observations, $t_{\rm UTC}=t_{\rm TAI}-37$ s, where TAI denotes International Atomic Time.

Circularized Gaussian fits to the supplied Rubin science-image PSF models give core FWHMs of $0.76''$, $0.93''$, $0.84''$, $0.97''$, and $0.90''$ in chronological order.

For comparison, MPEC~2026-Q95 reports $i=22.49$, $21.97$, $r=22.18$, $i=21.83$, and $r=21.84$ mag for the five Rubin epochs in chronological order \citep{McLeod2026N2MPEC}. The archived Horizons catalog law, $m_1=17.1+5\log_{10}\Delta+10\log_{10}r_h$ with distances in au, predicts 21.90, 21.82, 21.80, 21.80, and 21.78 mag at these epochs and 22.47 mag at Keck. The catalog law predicts total magnitudes without matching our passbands or apertures.

\subsection{Keck/LRIS reduction and calibration}\label{app:keck}
We reduced the LRIS images with \texttt{LPipe} \citep{Perley2019} and measured $g$ and $i$ from the bias- and flat-corrected frames before background-source removal. We followed the catalog calibration of \citet{Bolin2020CD3,Bolin2023NT,Bolin2025KY26}. We calibrate Keck $g$ and $i$ to the SDSS system using transformed Pan-STARRS1 reference-star photometry \citep{Tonry2012}, and the red measurement to Cousins $R_C$ on the Vega system. The $i$ calibration uses 28 independent G-like standards, per-frame zero points, and shared spatial terms. We measure $i$ in the native unregistered images and combine the calibrated linear fluxes with exposure weights. The $R_C$ zero point uses 101 stars measured in $3''$ apertures in shorter exposures taken immediately before the comet frames, assuming stable throughput after exposure normalization.

A trail-convolved Moffat fit to Keck $i$ field stars gives a model-dependent de-trailed width of approximately $0.61''$. The PSF widths come from separate sidereal reference frames and need not equal those in each comet exposure.

For the radial profiles in Figure~\ref{fig:psf}, we measure the flux in $0.3''$ annuli, subtract a clipped-median sky measured at $4''$--$6''$, and normalize each profile to the central annulus. For the sunward width, we use the same sky offsets and extract a $0.81''$-wide profile along the projected solar direction ($259.28^\circ$) from the unsmoothed sum of $R$ frames 53--54. Ten stars in sidereal frame 52 provide the reference PSF.

The signal exceeds the empirical background by at least $3\sigma$ through the $2.1''$--$2.4''$ annulus in $g$ and the $1.5''$--$1.8''$ annulus in $R$; the next $R$ annulus has a significance of $2.9\sigma$. Beyond $2.4''$ in $g$ and $2.1''$ in $R$, all bins lie within $1.6\sigma$ and $0.9\sigma$ of the background, respectively. These thresholds depend on sky subtraction and do not define a physical coma edge.

\subsection{Aperture photometry, uncertainties, and relative reflectance}\label{app:phot}
For the Keck color comparison we invert the Lupton stellar transformation,\footnote{\url{https://www.sdss4.org/dr17/algorithms/sdssUBVRITransform/}}
\begin{equation}
r=\frac{R_C+0.1837g+0.0971}{1.1837}.
\label{eq:rtransform}
\end{equation}
We propagate the independent native $g$, $R_C$, and $i$ errors through the transformation, retaining $\operatorname{cov}(g,r)$ and the color covariance because $r$ is derived from $g$ and $R_C$. The formal errors exclude additional uncertainty in applying a stellar transformation to the comet.

We use exact fractional-pixel aperture areas and clipped-median sky estimates from annuli of $4''$--$6''$ at Keck and $2.2''$--$2.8''$ in the Rubin cutouts. The Keck flux errors include blank-aperture scatter, source shot noise, and calibration uncertainties from resampling independent stars. We combine paired $i$-band blank measurements before clipping to retain the background covariance between frames. Rubin errors include variance-plane and median-sky terms but omit calibration and interpixel covariance.

Trial Keck centroid shifts of $\pm0.1$ pixel change the small-aperture magnitudes by at most 0.013, 0.004, and 0.014 mag in $g$, $R$, and $i$, respectively; Rubin shifts of $0.10''$ produce changes of 0.01--0.07 mag. The shifts are imposed offsets, not measured centroid errors.

We use passband-specific absolute solar magnitudes based on \citet{Willmer2018}. Rubin values are integrated through its response curves; Keck uses the corresponding standard-system values, including $M_{\odot,r}=4.65$ AB mag for derived SDSS $r$ and $M_{\odot,R_C}=4.43$ Vega mag for native $R_C$.

We adopt SDSS solar colors $(g-r)_\odot=0.44$ and $(r-i)_\odot=0.11$ mag.\footnote{\url{https://www.sdss4.org/dr17/algorithms/ugrizVegaSun/}} Solar-subtracted colors give relative reflectances $Q_b=10^{-0.4[(m_b-m_r)-(m_{\odot,b}-m_{\odot,r})]}$, which we display at nominal wavelengths 477, 623, and 763 nm. We normalize the reflectances by the linearly interpolated value at 550 nm; this normalization adds no independent datum. Comparisons with the 24 fixed Bus--DeMeo mean spectra \citep{DeMeo2009} use the two correlated colors and their full covariance, giving two degrees of freedom. Monte Carlo draws of the native magnitudes propagate the correlated reflectance errors.

The envelopes show the published pointwise $\pm1\sigma$ dispersion within the B, C, Ch, X, S, and D classes. We divide each mean and its dispersion by the class mean at 550 nm. We compare the independent July 9 LSST $r-i$ measurement using its native solar color of 0.112 mag \citep{Chandler2026}; only Keck colors enter the spectrum and template fits. Solar-color offsets and an alternative $R_C$-to-$r$ transformation can change the ranking of the neutral classes. Integrating through the SDSS responses retains B as the closest mean but requires extrapolation below the templates' 450 nm limit. The fit statistics exclude class scatter, solar-reference uncertainty, differential seeing, and comet-specific transformation errors.

\subsection{Gemini North/GNIRS astrometry and compact-source photometry}\label{app:gnirs}
We processed the November 3 GNIRS short-camera frames with the facility bad-pixel mask, nonlinearity correction, gain and noise propagation, and contemporaneous lamp-on/lamp-off flats. No dark matched the science configuration (two 30~s coadds); the standard had a matching dark. To remove the changing twilight sky and stable detector pattern, we fit the background at each pixel using exposures at other dither positions. We masked the crossing field star within 18 native pixels and estimated the skies separately for alternating subsets A and B before combining them.

The November 3 headers lacked a usable celestial WCS. We restored the nominal dither geometry with the supplied GNIRS patch and the DRAGONS 4.2.2 preparation step \citep{Labrie2023DRAGONS}. At repeated dither positions, the bright source moved as expected for a fixed star during target tracking: the measured east and north drift amplitudes were 48.35 and 11.87 arcsec h$^{-1}$, compared with 48.34 and 12.05 arcsec h$^{-1}$ predicted from the tracking rate. Gaia Data Release 3 source 6853295184686981248 lay $4.9''$ from the header-based position estimate; the next Gaia source lay $24.3''$ away. Its red Pan-STARRS1 counterpart (object 82383051031762254) has 64 catalog detections. The motion identifies a field star; the pointing and catalog match identify the adopted counterpart.

We fit the relative stellar positions in nine exposures across the commanded dithers, accounting for the star's drift. These displacements constrain the detector scale ($0.151$--$0.152''$ per pixel) and orientation; the catalog position sets only the absolute origin. For the other six exposures we propagated the fitted stellar motion. Gaia gives no proper motion for this star, so we estimated its November 2026 position using Pan-STARRS1, Gaia, and NOIRLab Source Catalog positions spanning 2012--2019, inflating the errors for excess scatter. The catalog, centroid, and dither terms give WCS standard errors of $0.148''$ east and $0.135''$ north. We registered the frames using topocentric Horizons positions at the exposure midpoints. The stack coordinates refer to 04:49:37.050 UTC on November~3, the reference exposure. For this registration we used the November~3 Horizons solution~2; Section~\ref{sec:orbit} identifies the August solution used for the orbital integrations. Adding the independent Horizons and WCS covariance matrices gives $3\sigma$ semiaxes of $2.54''$ and $0.48''$ at position angle $78.2^\circ$ east of north, conditional on the stellar identification and plate solution. The feature lies 0.66 standard deviations from the prediction in the metric of this covariance. No second catalog star independently checks the absolute solution; the ellipse tests positional consistency and does not give the probability that the feature is N2.

We fit a circular Gaussian ($\sigma=2.1$ native pixels) and a local background plane to the 11 individual sky-subtracted, unsmoothed frames with unmasked coverage at the fixed candidate position. We estimated the flux error from 622 overlapping blank placements to account for correlated sky noise. Transparency changes and a mismatch between the source and the adopted profile can affect the compact-flux estimate.

We calibrated the GNIRS $J$-band images with five observations of SA~110$-$361 obtained on UT 2026 November 3. We transformed the standard's Two Micron All Sky Survey (2MASS) photometry to the Mauna Kea Observatories (MKO) system \citep{Leggett2006MKO} and adopted $J_{\rm MKO,Vega}=11.188$ mag. The five count rates measured in a $2.7''$-radius aperture had a root-mean-square scatter of 1\%. We used the observed growth curve to correct these rates to a $4.2''$ reference radius. With $k_J=0.05$ mag per airmass, we transferred the zeropoint from the standard's mean airmass of 1.53 to N2's mean airmass of 1.39, obtaining $ZP_J=26.972$ mag for count rates in electrons s$^{-1}$. The $\simeq0.06$ mag calibration floor primarily reflects uncertainty in the standard's catalog magnitude, the 2MASS-to-MKO transformation, and the aperture correction.

On November 20 we reduced the two filters separately with DRAGONS. We applied the bad-pixel mask, nonlinearity and gain corrections, and normalized flats from eight lamp-on and eight lamp-off exposures per filter. No dark matched either science sequence. We masked sources and estimated each exposure's sky from other dithers in the same filter, fitting the changing $J$ twilight background per detector pixel. The header WCS gave the scale and orientation. We refined its translation using four Legacy Imaging Surveys Data Release 10 objects in $J$ and five in $Y$, allowing the $Y$ correction to vary with time.\footnote{\url{https://www.legacysurvey.org/dr10/description/}} We added the field-registration covariance to that of the topocentric Horizons solution~3 at Maunakea. The combined $3\sigma$ semiaxes are $2.37''\times0.64''$ in $J$ and $2.20''\times0.64''$ in $Y$.

We adopted $J_{\rm MKO,Vega}=11.434\pm0.014$ mag and $Y_{\rm Vega}=11.697\pm0.009$ mag for S813-D from the United Kingdom Infrared Telescope standards catalog\footnote{\url{https://about.ifa.hawaii.edu/ukirt/text-file-standard-stars/}} and \citet{Hillenbrand2002Y}, respectively. The standard-star count rates include the measured growth correction to a $4.2''$ reference radius. With extinction of 0.05 mag per airmass, the zeropoints at mean science airmasses 1.42 ($J$) and 1.53 ($Y$) are 26.9108 and 26.6130 mag for rates in electrons s$^{-1}$. The calibration floors include the standard magnitudes, frame scatter, aperture and background terms, and extinction transfer. A single standard does not determine the color term between GNIRS $Y$ and the Hillenbrand passband for N2.

At the fixed $Y$ offset and separately at the exact ephemeris position, we fit the standard-star Moffat profile integrated over native pixels with a local sky plane on each unsmoothed exposure. All 17 $J$ and eight of the ten $Y$ exposures had adequate unmasked coverage for the fit at the selected position. We estimated the pointwise errors at off-target positions of comparable coverage. For the apparent color we used the unrounded count-rate ratio $R=F_Y/F_J$, so $Y-J=ZP_Y-ZP_J-2.5\log_{10}R$. We transformed the 68.27\% Fieller interval for $R$ to magnitudes, treating the two empirical flux errors as independent Gaussian standard deviations conditional on the selected position.

\begin{figure}[!htbp]
\centering
\includegraphics[width=0.94\textwidth]{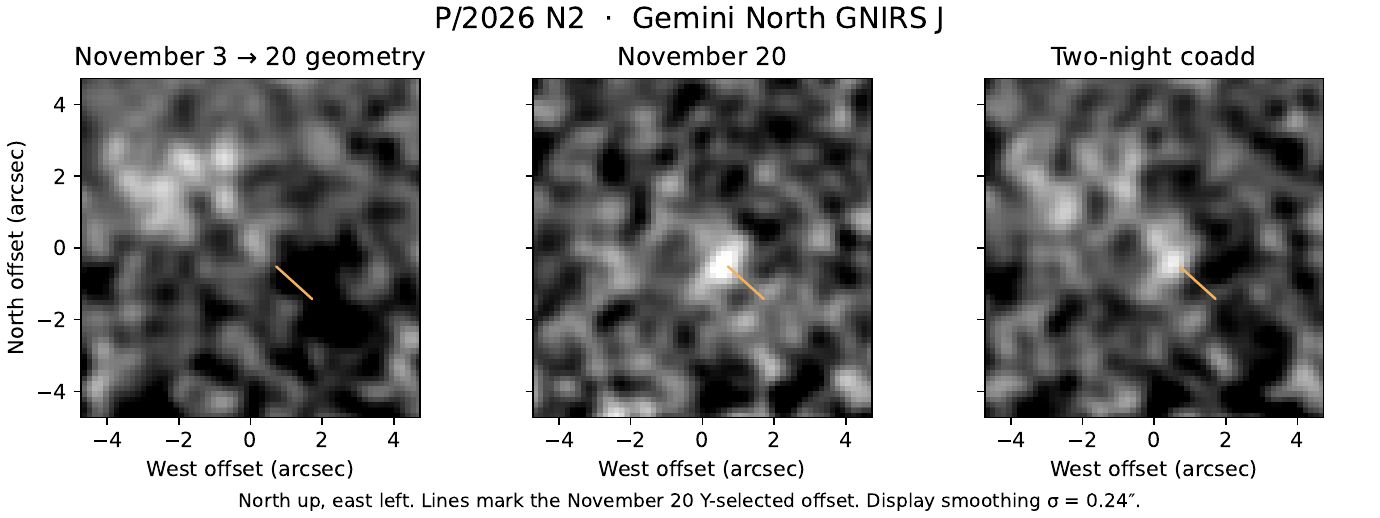}
\caption{Gemini North/GNIRS $J$-band persistence check for P/2026 N2. From left to right: November 3 (15 exposures, 900~s), November 20 (17 exposures, 1,020~s), and their weighted comet-frame mean (32 exposures, 1,920~s nominal). We scaled November 3 to the November 20 distances and phase angle under constant intrinsic activity before combining the images. Each origin is the Horizons solution~3 prediction; north is up and east is left. The orange lines point to the offset selected in November~20 $Y$ ($0.45''$ west, $0.30''$ south). Each panel is stretched separately and smoothed with a Gaussian of $\sigma=0.24''$ for display. A $0.60''$-radius aperture on the unsmoothed images at this offset gives pointwise S/N of 1.31, 4.27, and 3.86, respectively. At the predicted position the coadd gives S/N 2.97.}\label{fig:gnirscoadd}
\end{figure}

For the two-night $J$ comparison (Figure~\ref{fig:gnirscoadd}), we aligned the 15 November 3 exposures (the A+B stack) and 17 November 20 exposures on the topocentric solution~3 predictions at the stack reference epochs, November 3 04:49:37.050 and November 20 04:42:26.800 UTC. The November 3 origin already includes the catalog-star proper-motion correction.

We modeled constant activity as an unchanged effective $J$-band scattering cross section for the compact source. Using Equation~\ref{eq:h}, we scaled its November 3 flux to November 20 by $g_{3\to20}=[(r_{h,3}\Delta_3)/(r_{h,20}\Delta_{20})]^2\,\Phi(\alpha_{20})/\Phi(\alpha_3)$, with subscripts giving the November dates. We used the Maunakea geometry at the exposure-weighted $J$ epoch on each night. As $\alpha$ decreased from $23.8450^\circ$ to $22.5385^\circ$, the Halley--Marcus phase function increased from 0.4675 to 0.4816 (flux factor 1.0301). The distance factor was 0.8362, giving $g_{3\to20}=0.8613$, or $+0.162$ mag. This assumes the dust phase function applies in $J$; its uncertainty is not included in the statistical errors. The $0.60''$ aperture spans projected radii of 1,027 and 1,120 km on the two nights, respectively. The compact-source scaling assumes comparable enclosed fractions and is not a correction for a resolved coma.

We multiplied the unsmoothed November 3 rates by $g_{3\to20}$ and by the independent zeropoint factor 0.9454 (combined rate factor 0.8143), and the variances by the square of that product. We projected the result onto the November 20 $0.15''$ grid, propagating the masks, coverage, pixel-area ratio (0.9779), and diagonal variances through bilinear interpolation. We averaged rates using per-pixel inverse-variance weights scaled to the blank-aperture scatter on each night. The 32 distinct exposures total 1,920~s nominally; the mean exposures within the selected aperture are 546 and 950~s from the two nights.

We measured enclosed light with a $0.60''$-radius aperture and median sky in a $1.5''$--$2.4''$ annulus on each regridded $J$ image and their coadd. Off-target apertures of comparable coverage and propagated noise give the empirical errors, including contributions from resampling and shared skies. For the mixed-epoch color we measured the November 20 $Y$ stack with the same aperture and annulus, without a seeing or aperture correction. The reported color interval excludes uncertainty in the nightly calibration transfer and adopted phase function. The November 20 $J$ and $Y$ stack reference epochs are 31 min apart ($Y$: 05:13:48.300 UTC); Table~\ref{tab:photo} gives exposure-weighted means.

\subsection{CFHT archival image search}\label{app:cfht}
The search with Horizons solution~2 osculating elements and a $5'$ margin around the predicted track returned 16 distinct CFH12K exposures. We evaluated the topocentric position at Maunakea at each exposure midpoint and used the image astrometry to identify usable detector pixels. On September 24 the nominal position fell outside the valid area in two CFH12K exposures and near the edge in a third, truncating the source aperture.

For September 23 we averaged the four sky-subtracted images (360, 360, 480, and 360~s) in the comet frame after normalizing by exposure time. We measured the flux in a $1.5''$-radius aperture with a $3''$--$6''$ sky annulus. We estimated $\sigma$ as 1.4826 times the median absolute deviation of fluxes in fully covered blank apertures. The formal $3\sigma$ ephemeris ellipse had semiaxes of $24.2''$ and $0.67''$. Registering the images from both nights on field stars showed that the off-center feature remained at the same sky position while the corresponding position along N2's predicted track moved about $19''$ away.

\FloatBarrier
\setcounter{equation}{0}
\section{Dust production and retention models}\label{app:dustmodels}
\subsection{Conversion of aperture flux to mass-loss rate}\label{app:mass}
In the single-scattering model of \citet{Fink2012}, $A$ in $Af\rho$ depends on scattering efficiency and angle and is distinct from geometric albedo $p_V$. We assume an optically thin, steady, isotropic coma of identical spherical grains with radius $a_d$, density $\rho_d$, and constant radial speed $v_e$. For a production rate $Q_N$ grains s$^{-1}$, the number density at nucleus-centered distance $r$ is
\begin{equation}
n(r)=\frac{Q_N}{4\pi v_e r^2}.
\label{eq:dust_density}
\end{equation}
Integrating along the line of sight $z$ and over a circular aperture of projected radius $\rho$ gives
\begin{equation}
\begin{aligned}
N_{\rm ap} &=\int_0^\rho 2\pi b\,db\int_{-\infty}^{\infty}
\frac{Q_N\,dz}{4\pi v_e(b^2+z^2)}\\
&=\frac{\pi Q_N\rho}{2v_e},
\end{aligned}
\label{eq:dust_aperture_number}
\end{equation}
where $b$ is projected distance from the nucleus. The factor $\pi/2$ arises because the aperture samples a cylinder through the spherical coma, giving an effective residence time $N_{\rm ap}/Q_N=\pi\rho/(2v_e)$. The total geometrical cross section is $C=\pi a_d^2N_{\rm ap}$, and the filling factor is $f=C/(\pi\rho^2)$.

Using the geometric-albedo scattering convention \citep{Hsieh2009Read}, the comet and solar fluxes satisfy $F_c/F_\odot=p_V\Phi(\alpha)C/(\pi r_h^2\Delta^2)$, with $F_\odot$ at 1 au in the same band and $\Phi(0)=1$. Here $r_h$ is in au, and $C$, $\Delta$, and $\rho$ use consistent length units. Substitution into the photometric definition $Af\rho=4r_h^2\Delta^2(F_c/F_\odot)/\rho$ gives
\begin{equation}
A(0^\circ)f\rho=\frac{Af\rho}{\Phi(\alpha)}
=\frac{4p_VC}{\pi\rho}.
\label{eq:dust_cross_section}
\end{equation}
Thus $A(0^\circ)=4p_V$ under this convention. Multiplying $Q_N$ by the mass per grain, $(4\pi/3)\rho_da_d^3$, gives Equation~\ref{eq:dust_rate}.

\subsection{Cross section remaining after an impulsive release}\label{app:retain}
We propagate the Keck-epoch Horizons state from Section~\ref{sec:fp} backward to the June 27 release time $t_0$. At release the grain and nucleus have the same heliocentric position and velocity. Thereafter their position vectors $\mathbf r_d$ and $\mathbf r_N$ obey
\begin{equation}
\ddot{\mathbf r}_N=-\mu_\odot\frac{\mathbf r_N}{|\mathbf r_N|^3},\qquad
\ddot{\mathbf r}_d=-\mu_\odot(1-\beta)\frac{\mathbf r_d}{|\mathbf r_d|^3},\qquad
\beta=\frac{B}{a_d}.
\label{eq:impulse_motion}
\end{equation}
Here $\mu_\odot=GM_\odot$, $a_d$ is the grain radius, and we adopt $B=1~\mu{\rm m}$. For each exposure, $\hat{\mathbf n}$ points from the observer toward N2 and $\Delta$ is its distance. We use the LSST cutout coordinates and the Keck observing geometry to set the respective sightlines. Let $\mathbf u$ point from the observer at reception to the grain at emission, and let $\psi$ be its angular separation from N2. The grain's line-of-sight and transverse offsets from the observer give the projected distance in the plane through N2:
\begin{equation}
D_{\rm sky}(t,\beta)=\Delta\tan\psi
=\Delta\,\frac{|\mathbf u-(\mathbf u\!\cdot\!\hat{\mathbf n})\hat{\mathbf n}|}
{\mathbf u\!\cdot\!\hat{\mathbf n}}.
\label{eq:impulse_projection}
\end{equation}
We evaluate the nucleus at $t-\Delta/c$ and the grain at $t-|\mathbf u|/c$, where $t$ is the reception time and $c$ is the speed of light. Projecting onto the sightline at each epoch accounts for the changing viewing geometry across opposition. For projected aperture radius $\rho$, the radiation-pressure ratio $\beta_c$ at the edge sets the limiting grain radius $a_c$:
\begin{equation}
D_{\rm sky}(t,\beta_c)=\rho=2{,}000~{\rm km},\qquad a_c(t)=B/\beta_c.
\label{eq:impulse_cutoff}
\end{equation}
At the July 9 and 14 $r$ exposures and the October 7 $R_C$ stack, the elapsed times since release are 11.91, 16.92, and 101.91 days. The respective roots are $\beta_c=0.03381$, $0.02127$, and $0.0005732$. With $B=1~\mu{\rm m}$, these give $a_c=29.6~\mu{\rm m}$, $47.0~\mu{\rm m}$, and $1.745~{\rm mm}$. Larger grains have smaller $\beta$ and remain inside the aperture at these epochs. The limiting radii depend on the assumed release time, ejection speed, and $B$.

For a differential size distribution $dN/da_d=\Gamma a_d^{-s}$ on $[a_{\min},a_{\max}]$, grains of radius $a_d$ each contribute area $\pi a_d^2$. Writing $a_\ell(t)=\max[a_{\min},a_c(t)]$, we integrate over the sizes still in the aperture:
\begin{equation}
C_d(t)=\pi\Gamma\int_{a_\ell(t)}^{a_{\max}}a_d^{2-s}\,da_d
=\pi\Gamma\begin{cases}
\big[a_{\max}^{3-s}-a_\ell(t)^{3-s}\big]/(3-s), & s\ne3,\\
\ln[a_{\max}/a_\ell(t)], & s=3.
\end{cases}
\label{eq:retained_dust}
\end{equation}
Here $\Gamma$ is the normalization and $C_d=0$ if $a_\ell\geq a_{\max}$. The asymptotic scaling $C_d\propto(t-t_0)^{6-2s}$ assumes constant projected acceleration, $s>3$, and $a_{\min}\ll a_c\ll a_{\max}$; constant brightness alone does not imply $s=3$.

If dust supplies all the light and its scattering properties are fixed, the corrected flux in the same aperture is proportional to $C_d$. We compare October with the inverse-variance-weighted mean of the two July magnitudes (Table~\ref{tab:largeap}). Since magnitudes are logarithmic, the corresponding July cross section is a weighted geometric mean. With normalized weights $w_9=0.288$ and $w_{14}=0.712$ from the July uncertainties of 0.11 and 0.07~mag, the model gives
\begin{equation}
f_{\rm ret}=\frac{C_d(t_{\rm Oct\,7})}
{C_d(t_{\rm Jul\,9})^{w_9}C_d(t_{\rm Jul\,14})^{w_{14}}}
=10^{-0.4\Delta H_r}.
\label{eq:impulse_fraction}
\end{equation}
For the measured difference, $\Delta H_r=-0.10\pm0.09$~mag, the central October-to-July flux ratio is $10^{-0.4(-0.10)}=1.096$. At two formal uncertainties toward fading, $\Delta H_r=-0.10+2(0.09)=+0.08$~mag and $f_{\rm ret}=10^{-0.4(0.08)}=0.929$. The two-formal-error tolerance allows a 7\% loss of scattering area. The tolerance excludes the systematics in Section~\ref{sec:brightness}. For $a_{\min}=1~\mu$m, Equations~\eqref{eq:retained_dust} and \eqref{eq:impulse_fraction} meet this tolerance with $a_{\max}\gtrsim2.4$ cm for $s=2$, or $a_{\max}\gtrsim25$ cm for $s=2.5$. A broad $s=3.5$ distribution retains at most approximately 15\% of its July cross section, predicting about 2.0 mag of fading. The same slope can retain 93\% if $a_{\min}\gtrsim1.6$ mm and $a_{\max}=1$ cm. The changing projection, unknown release time and ejection speeds, possible nucleus light, and treatment of the PSF and sky prevent the photometry from determining a unique size distribution. The calculation does not fit the aperture growth or resolved morphology.

\section{Numerical integration of the orbital clones}\label{app:integrations}
We generate 5,000 massless clones from the six-dimensional Gaussian defined by the JPL nominal elements and full covariance in the native coordinates $(e,q,T_p,\Omega,\omega,i)$ at JD~2460616.5 TDB (Table~\ref{tab:orbit}), where $T_p$ is perihelion time. The adopted August 25 JPL solution spans 1,822 days (80 observations, 2021 August 4 to 2026 July 31). Our eight October 7 Keck I positions (MPC code T16; MPEC~2026-V22; \citealt{MPC2026V22}) extend the arc by 68 days (3.7\%) to 1,890 days. At the epoch of Table~\ref{tab:orbit}, their inclusion in the November 3 JPL solution changes $a$ by $1.5\times10^{-6}$ au, well within its formal uncertainty of $1.35\times10^{-5}$ au. Given this small shift, we expect the concentrated outer Main Belt core to persist; the clone fractions in Section~\ref{sec:orbit} use the August solution without refitting. Neither solution includes the unreported July 4 DECam astrometry.

\begin{figure}[!htbp]
\centering
\includegraphics[trim=0bp 18bp 0bp 11bp,clip,width=0.92\textwidth,height=0.70\textheight,keepaspectratio]{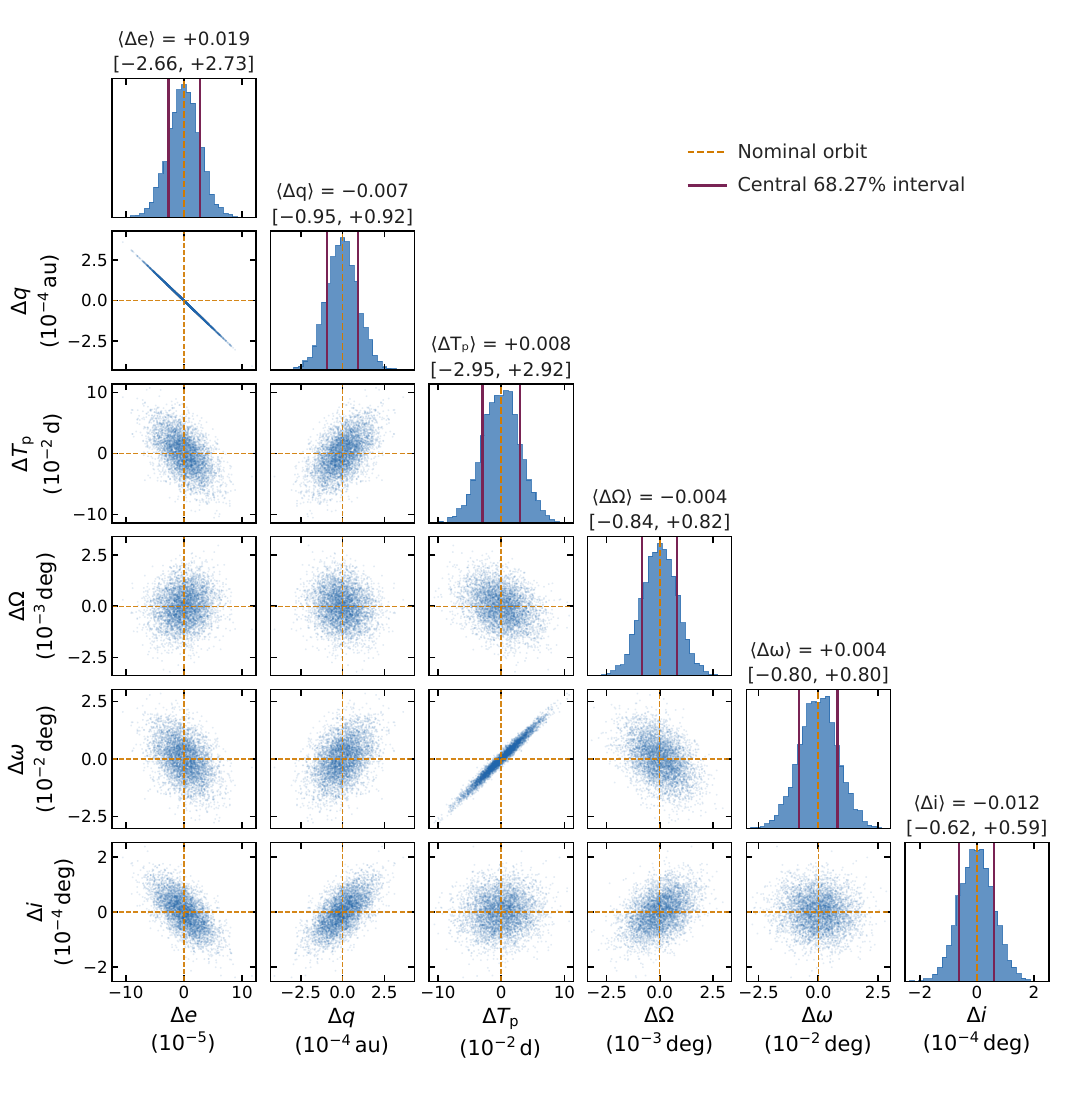}
\caption{Initial distribution of all 5,000 clones at JD 2460616.5 TDB. Diagonal panels show marginal histograms; lower-triangle panels show pairwise distributions in the native JPL coordinates $(e,q,T_p,\Omega,\omega,i)$. Axes show offsets from the nominal elements in the indicated units; orange dashed lines mark zero offset. Above each histogram we give the mean offset and the central 68.27\% interval in axis units. Brackets and solid plum lines mark the interval endpoints. The samples are reconstructed from the original full-covariance sampler and recorded seed. Their initial $a$, $e$, and $i$ summaries reproduce the archived ensemble to numerical precision.}\label{fig:clonecorner}
\end{figure}

We draw the six elements jointly: higher eccentricity is paired with smaller perihelion distance in the JPL fit ($R_{e,q}=-0.99994$). Independent draws would lose this correlation. Each clone has elements $\mathbf x=\mathbf x_0+D L\mathbf z$, where $\mathbf x_0$ is the nominal orbit and $\mathbf z$ contains six independent standard normal draws. The Cholesky factor $L$ is the lower triangular matrix satisfying $LL^{\mathsf T}=R$; it gives $L\mathbf z$ the JPL correlations, and the diagonal matrix $D$ scales the offsets by the published standard deviations. A fixed seed (20261009, NumPy~1.26.4) reproduces the 5,000 initial orbits in Figure~\ref{fig:clonecorner}. We retained all draws without rescaling the covariance.

We convert the sampled elements to barycentric Cartesian states and integrate them forward for 50~Myr with REBOUND~5.2.2 \citep{Rein2012REBOUND}, using the MERCURIUS hybrid integrator \citep{Rein2019MERCURIUS} with a 5-day base step. Close encounters use the adaptive 15th-order integrator IAS15 \citep{Rein2015IAS15}, with the default MERCURIUS encounter-switching parameters and safe mode retained. The Newtonian force model includes the Sun and the seven planetary systems from Venus through Neptune, combining the Earth--Moon mass. It omits Mercury, relativistic corrections, nongravitational accelerations, physical collisions, and particle removal. We record heliocentric osculating elements in the ecliptic J2000 frame every 10~kyr. In tests of 16 clones over $\pm10$~kyr, the semimajor axes and eccentricities agreed more closely with the IAS15 reference solution as we reduced the timestep from 5 to 2.5 and then to 1.25 days.
\FloatBarrier

\end{document}